\documentclass[%
aip,
pof,
preprint,
floatfix
]{revtex4-2}

\usepackage[dvipsnames]{xcolor}

\usepackage[a4paper,margin=2cm]{geometry}
\usepackage{graphicx}
\usepackage{subcaption}
\usepackage[dvipsnames]{xcolor}
\usepackage{tcolorbox}
\usepackage{enumitem}
\usepackage{float}
\usepackage{setspace}
\usepackage{url}
\usepackage{hyperref}

\usepackage{upmath}
\usepackage{amsmath,amssymb}
\usepackage{amsbsy}

\providecommand\bnabla{\boldsymbol{\nabla}}
\providecommand\bcdot{\boldsymbol{\cdot}}
\providecommand{\e}[1]{\ensuremath{\cdot 10^{#1}}}

\preprint{AIP/123-QED}
	
\begin{document}

\title{Diffusion of turbulence following both stable and unstable step stratification perturbations.}
% Force line breaks with \\

\author{L. Gallana}
% \email{federico.fraternale@uah.edu}
%\affiliation[Author's affiliation at the time of the present analyses: ]{Dipartimento di Ingegneria Aerospaziale, Politecnico di Torino, Torino, Italy 10129.}
\author{S. Abdunabiev}
% \email{federico.fraternale@uah.edu}
%\affiliation[Author's affiliation at the time of the present analyses: ]{Dipartimento di Ingegneria Aerospaziale, Politecnico di Torino, Torino, Italy 10129.}
\author{M. Golshan}%
% \email{gabrile.nastro@isae-supaero.fr}
%\affiliation{Dipartimento di Scienza Applicata e Tecnologia, Politecnico di Torino, Torino, Italy 10129.}%Lines break automatically or can be forced with \\
\author{D. Tordella}
\email[Author to whom correspondence should be addressed: ]{daniela.tordella@polito.it}
%  \homepage{http://www.Second.institution.edu/~Charlie.Author.}
\affiliation{
	Dipartimento di Scienza Applicata e Tecnologia, Politecnico di Torino, Torino, Italy 10129}%\\This line break forced% with \\
\date{\today}
\graphicspath{{figs/}}
\newcommand{\fr}{Fr$^2$}

\maketitle

\section*{Abstract}

\vspace{-0.2cm}  

{The evolution of a two-phase, air and unsaturated water vapor, time decaying, shearless, turbulent layer has been studied in the presence of both stable and unstable perturbations of the normal temperature lapse rate. The top interface between a warm vapor cloud and clear air in the absence of water droplets was considered as the reference dynamics.} Direct, 3D numerical simulations were performed within a 6m x 6m wide and 12m high cloud portion, which was hypothesized to be located close to an interface between the warm cloud and clear air. The Taylor micro-scale Reynolds' number was 250 inside the cloud portion. The squared Froude's number varied over intervals of [0.4; 1038.5] and [-4.2; -20.8]. A sufficiently intense stratification was observed to change the mixing dynamics. The formation of a sub-layer inside the shearless layer was observed. The sub-layer, under a stable thermal stratification condition, behaved like a pit of kinetic energy. On the other hand, it was observed that kinetic energy transient growth took place under unstable conditions, which led to the formation of an energy peak just below the center of the shearless layer. The scaling law of the energy time variation inside the interface region was quantified: this is an algebraic law with an exponent that depends on the perturbation stratification intensity. The presence of an unstable stratification increased the differences in statistical behavior among the longitudinal velocity derivatives, compared with the unstratified case. Since the mixing process is suppressed in stable cases, small-scale anisotropy is also supressed.

KEYWORDS: Turbulent transport, thermal stratification, stability, initial value problem, passive scalar.
	
\section{Introduction}

Warm clouds, such as stratocumuli, swathe a significant part of the earth's surface and play a major role in the global dynamics of the atmosphere by reflecting incoming solar radiation - thus contributing to the Earth's albedo - so that an accurate representation of their dynamics is important for the large-scale analyses of atmospheric flows \cite{wo12}. Their dynamics are controlled by the close interplay that takes place among radiative driving, turbulence, surface fluxes, latent heat release, entrainment,
and the energy captured from acoustic-gravity waves propagating into clouds from below or above cloud layers, or from cosmic rays during their interaction with water drops. The introduction of all these aspects into numerical simulations is still not the state of the art. For instance, compressibility should be included in a numerical simulation to account for internal acoustic and gravity waves and baroclinicity effects, but efficient techniques that are able to carry out the simulation of clouds at the relevant evanescent values of the Mach number have not yet been developed. However, among all these physical effects, turbulent mixing and entrainment-detrainment processes at the top of a cloud have been identified as being of fundamental importance to determine the internal structure of warm clouds, so that a clear and complete understanding of their physics can be obtained \cite{ger2013}.

Stratification in the atmosphere is usually stable above the boundary layer \cite{sheu}, i.e. a fluid particle that is displaced in the vertical direction tends to return to its initial position. However, unstable perturbations of local stratifications can be expected during the formation and disruption phases of clouds. Terrestrial rotation becomes of secondary importance in  local atmospheric dynamics, and the stratification effects dominate \citep{Vallis06,Gill1982}. Over the last few decades, there have been important advances in the understanding of turbulence in the presence of intense stratification. {For example, in the homogeneous stratified turbulence context, it is known that isotropic turbulence in a stratified fluid initially rapidly becomes anisotropic, with the formation of pancake-like structures on its inside \citep{lin79,kimura96}}.
As pointed out by Malinowski et al. 2013 \cite{mal2013}, data from most field campaigns and large-eddy simulations are too poorly resolved to infer the details of the interfacial layer,  even though it is known that a high level of turbulence must be present for entrainment to take place.
For this reason, in this work, we have studied transport across an  unsaturated vapor cloud - clear air interface through DNS (Direct Numerical Simulation). 

{While we have considered turbulent transport without shear in thermal stratification conditions, and have also included the Lagrangian dynamics of both monodisperse and polydisperse populations of water droplets in two recent works \cite{golshan2021,fossa2022}, we here focus on the phase preceding the formation of a warm cloud containing a liquid phase. We therefore focus on the turbulent transport of the unsaturated vapor phase, considered as a passive scalar, and on the associated temperature field, considered as an active scalar. This has allowed us to consider a better spatial resolution by adopting the two-dimensional stencil parallelization method.
%and, therefore, to clearly understand the details of the internal structure of clear air - vapor cloud interfaces. 
In fact, this parallelization technique of the three-dimensional DNS code cannot be efficiently adopted in the presence of discrete elements, such as water droplets transported in a Lagrangian way, because of a large latency in the communication among processes (cores). A numerical code for the study of the growth, collision, coalescence and clustering water droplets inside turbulent, warm, 
cloud-clear air interfaces is discussed here in detail \citep{ruggiero2020numerical}}

{Thus, we have focused on how the dynamics of the smallest scales of an air flow influence vapor and thermal turbulent transport. We have therefore simulated an idealized configuration to better understand, under controlled conditions, the basic phenomena that occur at the vapor cloud interface over length scales of the order of a few meters}. Under these conditions, we have solved scales from a few meters to a few millimeters, that is, we have resolved  only the small-scale part of the inertial range and the dissipative range of the power spectrum in a small portion ($6$ m by $ 6$ m by $12$ m) of the atmosphere  across a vapor cloud - clear air interface. This has allowed us to investigate the entrainment dynamics that occurs in a thin layer at the top of the cloud, which has a smaller scale than the scale explicitly resolved in the large eddy simulations of clouds \cite{moe2000}. In this preliminary work, we have focused on two concomitant aspects of the top mixing layer of a vapor cloud: the effect of the presence of stratification and that of a turbulent kinetic energy gradient.  We have not considered  wind shear or %the phenomena linked to evaporation, condensation or 
radiative cooling processes, which are important in the presence of buoyancy reversal \cite{mel2010, mel2014}. Therefore, our simulations have been  performed by applying the Boussinesq approximation to Navier-Stokes momentum and energy equations, together with an advective-diffusive passive scalar transport equation. Details on the considered physical problem we have considered and on the governing equations are given in section 2. Section 3 contains some of our main results pertaining to intermittency, energy redistribution and entrainment. The concluding remarks are given in Section 4.

\section{The physical problem}

\begin{figure}[bht!]
	\centering
	\begin{minipage}{.7\textwidth}
		\includegraphics[width=\textwidth]{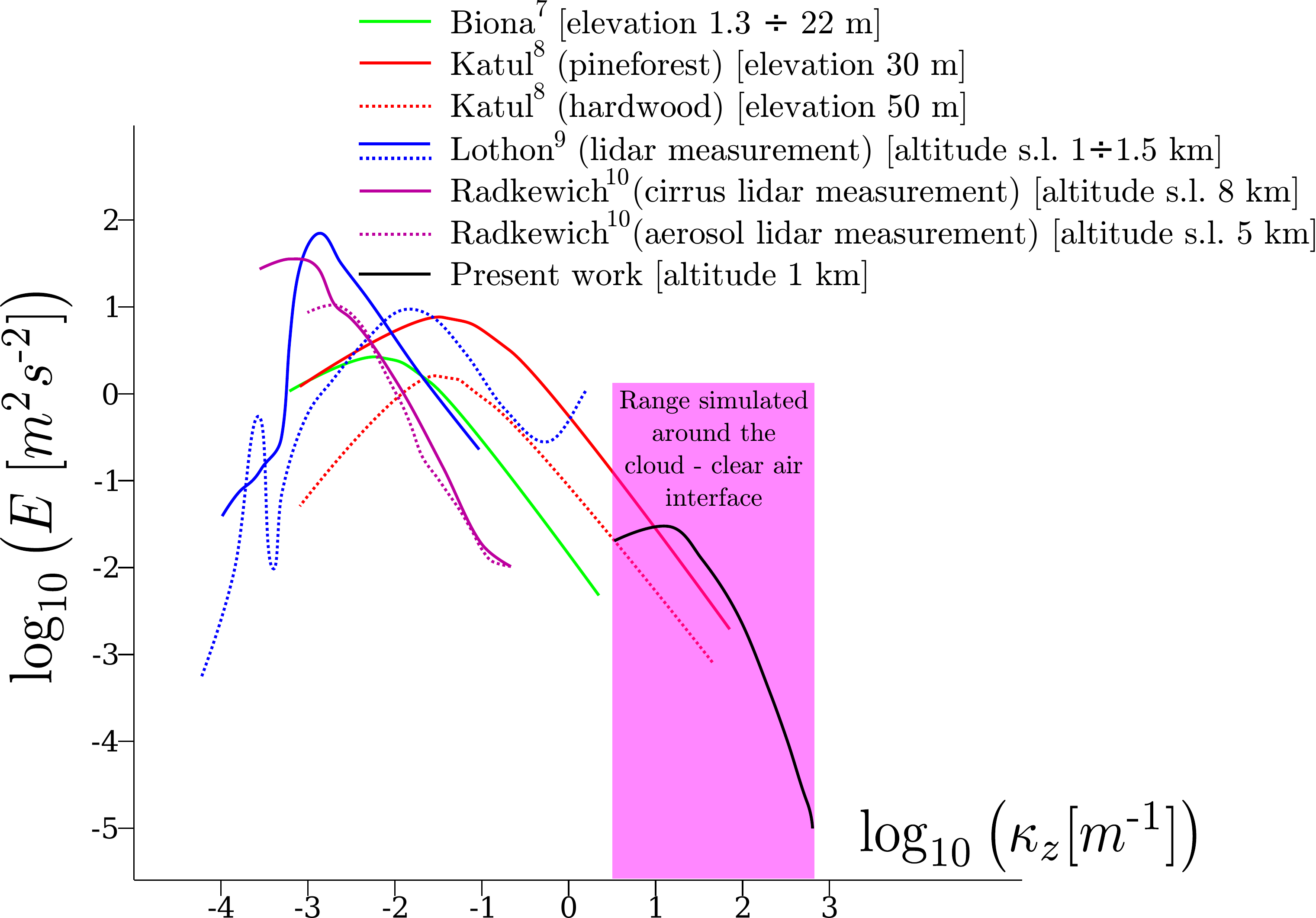}
	\end{minipage}\hfill
	\caption{Kinetic energy spectra. Contextualization of the present study (black spectrum,  inertial small-scale and dissipative ranges) to in-situ atmospheric measurements %\cite{biona, katul, lot, radk} 
	(colored spectra:  energy injection and low wave-number inertial scales). The aim of the current simulations is to represent the small-scale range of the spectrum that in situ measurements have not been able to detect.}
	\label{fig:spett}
\end{figure}

\begin{figure}[bht!]
	\centering
	\begin{minipage}{.7\textwidth}
	\includegraphics[width=\textwidth]{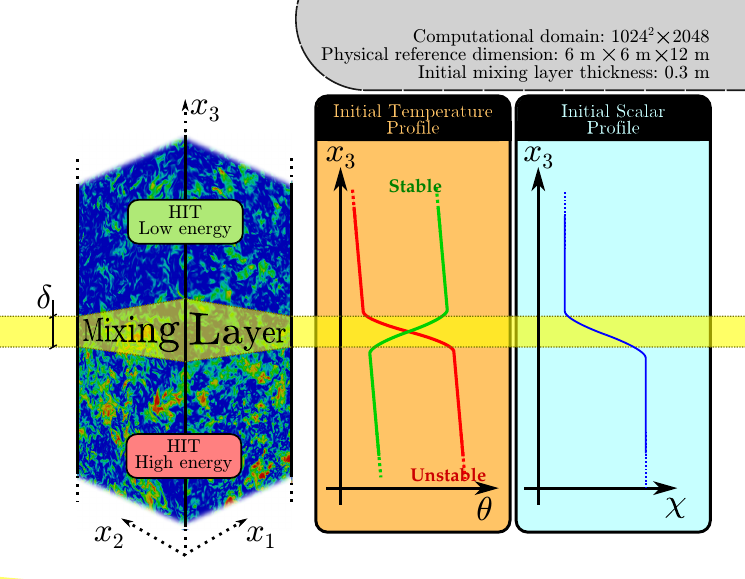}
		\end{minipage}\hfill
		\caption{Scheme of the initial conditions. $E_1$ is the mean initial turbulent kinetic energy below the shearless mixing layer (cloudy vapor, HIT high energy region), $E_2$ is the same, but for the top (clear air, HIT low energy region). We assume $E_1/E_2=6.7$ for this model of the top interface. The stratification inside this interfacial mixing is represented by a local temperature perturbation with respect to the neutral profile; the perturbation can be either stable or unstable. The unsaturated vapor (passive scalar) is initially only present in the cloudy high-energy region. Gravity is opposite to the positive $x_3$ direction.}
	\label{fig:schema}
\end{figure}

We considered the interaction of two {homogeneous isotropic turbulence air fields, with different levels of kinetic energy and unsaturated water vapor (passive scalar)}, in a $6$ m by $6$ m by $12$  m domain. As can be seen in Fig. \mbox{\ref{fig:spett}}, the chosen domain size allowed us to simulate the highest wave numbers of the spectral inertial range and the dissipative range of  {\em  in situ} measurements of the atmospheric power velocity spectra.  As shown in Fig. \mbox{\ref{fig:schema}}, the two HIT regions that make up the system interact through a shearless mixing layer, whose initial thickness was set to the same order of the integral scale as the air turbulence background $\ell$, which here has been assumed equal to $3\cdot10^{-1}$ m.  %{\color{red}Fig. \ref{fig:scalVIS}, show one example of visualization of the water vapor mixing ratio fluctuation within a stably stratified shear-less layer. COMMENT: CORRESPONDING FIGURE REMOVED}

The two isotropic regions (external to the mixing) have different kinetic energies. The underlying region is the more energetic one, and it is constituted by the vapor cloudy region. It hosts the passive scalar, which is our model for the water vapor phase, and has a kinetic energy equal to $E_1=0.06\ m^2/s^2$; the root mean square of the velocity in this region is $u_{rms}=0.2 $ m/s. The initial Taylor microscale Reynolds number, Re$_{\lambda} $, is approximately equal to $250$ ($\lambda$ is the Taylor scale).  The kinetic energy ratio between the two regions is equal to 6.7. This energy ratio is of the same order as the ones measured in warm clouds (see, e.g. \cite{mal2013}) and, furthermore, it allows our results to be compared with laboratory and numerical experiments on turbulent shearless mixing (see \cite{vw89, prl11}) in the absence of any stratification.

Buoyancy is taken into account through  perturbation, {$\theta'$, of the profile of the temperature distribution, $\theta$}, inside the troposphere, which is located across the shearfree mixing layer. The Prandtl number considered here is Pr$=0.74$ (standard atmosphere, altitude of $1000 \ $ m s.l.). The initial conditions for the temperature perturbation are  described in Figure \ref{fig:schema} and in Table \ref{tabone}. The ratio between the inertial and buoyancy forces is expressed by the Froude number Fr, which is defined as 

%\begin{equation}

	$$ \mathrm{Fr }=\frac{u_{rms}}{\ell \mathcal{N}}, \ \ \ \ \ \mathcal{N}^2 = {\alpha g \frac{\mathrm{d}\theta}{\mathrm{d} x_3 } }\ \ \	\ \ \ $$
		\
%\end{equation}

\noindent where $u_{rms}$ is the root mean square of the velocity fluctuation at the lower border of the interfacial layer, $\ell$ is the macroscale length inside the cloudy region, $\mathcal{N}$ is the Brunt-V\"ais\"al\"a\ frequency, $\theta$ is the mean temperature, $g$ is the gravitational acceleration and  $\alpha$ is the thermal expansion coefficient. We consider the square of the Froude number, \fr , which is based on the maximum gradient within the initial interface, to characterize each simulation. The initial values of \fr\ range from 1038.5 (negligible stratification) to 0.4 (strong stable stratification). It should be noted that our usage of \fr , instead of Fr, is due to the fact that we consider unstable cases. In fact, in such situations, $\mathcal{N}^2$ is negative for the initial temperature gradient -- and the Brunt-V\"ais\"al\"a\ frequency is imaginary; it actually yields the amplification rate of the perturbations. The most unstable stratification we therefore consider has a \fr\ equal to -4.2.

The unsaturated water vapor is taken into account by considering its normalized concentration $\chi$, which is equal to 1 in the lower cloudy region and to 0 in the upper clear-air region. Water vapor is considered as a passive scalar, with a Schmidt number Sc=0.61 (standard atmosphere, altitude of $1000\ m$ s.l.).

%\begin{figure}
%	\centering
%	\begin{minipage}{.8\textwidth}
%		\includegraphics[width=\textwidth]{500_u3_120}
%	\end{minipage}\hfill
%	\caption{\label{fig:scalVIS}{Visualization of the fluctuation of the vertical component of the velocity in the mixing layer between
%two isotropic turbulent flows in presence of a stable stratification.
%Vertical displacements indicate the module of the fluctuations, the color identifies the direction (dark red upwards fluctuations, blue
%downwards ones). The image is taken after 8 initial eddy turnover times when by means of an intense redistribution of kinetic
%energy a well of energy is generated inside the mixing.
%The two isotropic flows have a different level of turbulent
%kinetic energy: the flow in the region of negative $x_3$
%(bottom part of the mixing) has an energy 6.7 higher than
%the one above the mixing layer and has an initial Taylor
%micro-scale Reynolds number equal to 250. The initial \fr is equal to $4.2$.}}	
%\end{figure}

%\begin{table}[H]
%	\centering
%	\begin{tabular}{|c|c|c|}
%		\hline
%		Label & Froude number & Comment\\
%		\hline\hline

\begin{table}[bht!]
	\centering
	\begin{tabular}{|*{6}{r}|}
	\hline
		\def~{\hphantom{0}}
		$\nabla_z \theta_{ic}\ [Km^{-1}]$ & $\Delta \theta \ [K]$ & $\mathcal{N}_{ic}\ [s^{-1}]$  & Fr & \fr\ & {Re}$_b$ \\[3pt]\hline\hline
		1.3\e{-2}                 & 4.0\e{-3}               & 2.13\e{-2} & 45.57 & 1038.5 & 0.7 \\\hline
		2.0\e{-1}                 & 6.0\e{-2}               & 5.24\e{-2} & 8.32 & 69.2 & 10.9 \\\hline
		6.7\e{-1}                 & 2.0\e{-1}               & 1.50\e{-1} & 4.56 & 20.8 & 36.3 \\\hline
		3.3~\hphantom{$10^{-0}$}  & 1.0~\hphantom{$10^{-0}$} & 3.35\e{-1} & 2.04 & 4.2 & 181.7\\\hline
		3.3\e{1}\hphantom{$^{-}$} & 1.0\e{1}\hphantom{$^{-}$} & 1.06~\hphantom{$10^{-0}$} & 0.64 & 0.4 & 1817.2\\\hline
    	-6.7\e{-1}                 & -2.0\e{-1}               & /\hphantom{$10^{-0}$} &  /& -20.8 & -36.3 \\\hline
		-3.3~\hphantom{$10^{-0}$}  & -1.0~\hphantom{$10^{-0}$} &/\hphantom{$10^{-0}$}  & / & -4.2 &-181.7\\\hline\hline
	\end{tabular}
	\caption{Initial stratification level parameters.
	%		$G$ is the maximum gradient of $\theta$, expressed in terms of the standard troposphere {lapse rate} $G_0=0.0065\ Km^{-1}$
	$\mathcal{N}_{ic}=\sqrt{\alpha g \frac{\partial\theta}{\partial x_3}}$ is the characteristic Brunt-V\"ais\"al\"a\ frequency of the initial condition (suffix ic). The Froude number, Fr$=\displaystyle\frac{u_{{rms}}}{\mathcal{N}_{ic}\ell}$, and the Reynolds Buoyancy Number, Re$_b=\displaystyle\frac{\varepsilon \mathcal{N}_{ic}^{-2}}{\nu}$, offer an indication of the order of magnitude of the buoyancy forces, compared with the inertial terms {($\varepsilon$ is the initial energy dissipation rate, $\ell$ is the initial value of the spatial integral scale ($0.3$ m), and $\nu$ is the 
	kinematic viscosity of air)}.}
	\label{tabone}
\end{table}

We use the continuity, momentum and energy balance equations within the Boussinesq approximation, which holds for small temperature variations \cite{drazin}, while we use an advective-diffusive transport equation for water mixing ratio:
\begin{eqnarray}
	\label{eq:system}
	\label{eq:mass}
	\bnabla\cdot\boldsymbol u' & = & 0
	\\
	\label{eq:mom}
	\frac{\partial \boldsymbol{u}'}{\partial t} + \left(\boldsymbol{u}'\bcdot\bnabla\right)\boldsymbol{u}' & = & -\bnabla \frac{\tilde{p}}{\rho}+\nu \nabla^2\boldsymbol{u}' + \alpha\boldsymbol{g}\theta'
	\\
	\label{eq:ene}
	\frac{\partial \theta'}{\partial t} + \boldsymbol{u}'\bcdot\bnabla \theta'+u_3G & = & \kappa \nabla^2\theta'
	\\
	\label{eq:scal}
	\frac{\partial \chi}{\partial t} + \boldsymbol{u}'\bcdot\bnabla \chi & = & d_\chi \nabla^2\chi,
\end{eqnarray}

%\begin{equation}
%\end{equation}
% Flow density can be rewritten in terms of temperature as
% \begin{equation}
% \rho=\rho_0\left(1-\alpha\theta\right)
% \label{eq:theta}
% \end{equation}
\noindent where $\theta=\theta_0+\tilde{\theta}(x_3)+\theta'(\boldsymbol{x},t)$ is the temperature, which is composed of the reference constant temperature $\theta_0$ at a given altitude, of the static component $\tilde{\theta}(x_3)=G_0 x_3$, where $G_0$ is the standard lapse rate, and of the fluctuation $\theta'(\boldsymbol{x},t)$; 
moreover, $\tilde{p}=p+\alpha g x_3\left(\theta_0+G_0x_3/2\right)$ is the total hydrodynamic pressure ($p$ is the fluid dynamic pressure, $\alpha$ is the thermal expansion coefficient, and $g$ is the gravity acceleration); 
$u'$ is the velocity fluctuation; and $\chi$ is the vapor concentration of the air - water vapor mixture. The constants $\kappa$ and $d_\chi$ stand for the thermal and water vapor diffusivity, respectively. This is a very consolidated basic model that is often used as a representation of Eulerian equations for turbulent fields to which the liquid water component can be added as a Lagrangian set of N pointlike droplets\citep{ireland2012, kumar2014, gotoh2016, Gotzfried2017, gotoh2018, li2020, golshan2021}.
% where the hydrodynamic pressure $\tilde{p}$ is given by
% \begin{equation}
% \tilde{p}=p+\alpha g x_3\left(T_0+Gx_3/2\right)
%  \label{eq:ptilde}
% \end{equation}
% A scalar advective-diffusive equation is used in order to model the transport of the water vapor concentration $\chi$
%These equations are solved using a Fourier-Galerkin pseudospectral method with a fourth order Runge-Kutta time integration \cite{ict01}.

The initial condition for the velocity field is obtained by means of a linear matching of two different HIT fields, $u_1$ and $u_2$, which are randomly generated by respecting the physical solenoid condition, the required integral scale and the mean kinetic energy, see {\cite{ti06}}. The initial energy profile along direction $x_3$ is obtained by coupling the $u_1$ and $u_2$ fields, using equation (\ref{eq:velinit}).
As far as the scalars are concerned, in analogy with previous work\citep{jturb, golshan2021}, the initial conditions (constant along directions $x_1$ and $x_2$) are obtained from equations \ref{eq:tempinit}  and \ref{eq:vapinit} for the temperature and water vapor concentration, respectively. These equations are listed below:
\begin{eqnarray}
	\label{eq:initcond}
	\label{eq:velinit}
	\boldsymbol{u'}(\boldsymbol{x},t=0) & = & \boldsymbol{u_1}(\boldsymbol{x}) p_1(x_3)-\boldsymbol{u_2}(\boldsymbol{x})(1-p_1(x_3) )
	\\
	\label{eq:tempinit}
	{\theta}(\boldsymbol{x},t=0) & = & \Delta\theta p_2(x_3) 
	\\
	\label{eq:vapinit}
	{\chi}(\boldsymbol{x},t=0) & = & p_1(x_3), 
\end{eqnarray}
where $u_1$ and $u_2$ are the two external HIT, $\Delta\theta$ is the initial temperature step, while the weight functions $p_1(x_3)$ and $p_2(x_3)$ are defined as:
\begin{eqnarray}
	\label{eq:peso}
	\label{eq:p1}
	p_1(x) &= &\frac{1}{2}\left[1+\tanh\left(a\frac{x_3}{L_3}\right)+\tanh\left(a\frac{x_3-L3/2}{L_3}\right)+\tanh\left(a\frac{x_3-L_3}{L_3}\right) \right]
	\\
	\label{eq:p2}
	p_2(x) & = & \frac{x_3}{L_3}-\frac{1}{2}\left[1+\tanh\left(a\frac{x_3-L_3/2}{L_3}\right) \right].
\end{eqnarray}

\begin{table}[bht!]
	\centering
	\begin{tabular}{|c|c|c|}
		\hline
		Label & Froude number & Condition\\
		\hline\hline
		2G$_0$ & $Fr^2 = 1038.5$ & Stable \\ \hline
		30G$_0$ & $Fr^2 = 69.2$ & Stable \\ \hline
		100G$_0$ & $Fr^2 = 20.8$ & Stable \\ \hline
		500G$_0$ & $Fr^2 = 4.2$ & Stable \\ \hline
		5000G$_0$ & $Fr^2 = 0.4$ & Stable \\ \hline
		-100G$_0$ & $Fr^2 = -20.8$ & Unstable \\ \hline
		-500G$_0$ & $Fr^2 = -4.2$ & Unstable \\	\hline
	\end{tabular}
	\caption{Conversion of the temperature gradient values,  expressed in terms of G$_0=0.0065 \; ^o$C/m, that is, the  standard lapse rate of the atmosphere, to \fr numbers.}
\end{table}

The simulations were performed using our in-house computational Navier-Stokes code, which implements a pseudo-spectral Fourier-Galerkin spatial discretization and an explicit low storage fourth-order Runge-Kutta time integration scheme. Evaluation of the non-linear (advective) terms is performed by means of the 3/2 de-aliased method \mbox{\cite{ict01}}.
%	For details, see also the section Software (Incompressible Turbulent Flows) in the web pages  \it{ www.polito.philofluid.it}. 
	\normalfont
The grid has $N\times N\times N_3$ points, with $N=2^{10}$ and $N_3=2 \; N$), for a total of $2^{31}$ grid-points. Such a grid allows us to capture all the turbulent scales from the largest (integral scale $\ell$) to the smallest (Kolmogorov scale $\eta$). 
In fact, it should be noted that since the turbulence intensity, and thus the dissipation rate, decay in time, the small scales, in particular the Kolmogorov scale, $\eta_k$, grow in time. The grid size of 5.86 mm inside the mixing region matches the $k_{max} \eta \sim 3$ requirement for about two eddy turnover times. %, where the turbulent energy intensity falls somewhere between the intensities inside external homogeneous regions,

\begin{figure}[bht!]
	\centering
	    \begin{subfigure}[t]{0.48\textwidth}
			\includegraphics[width=\textwidth]{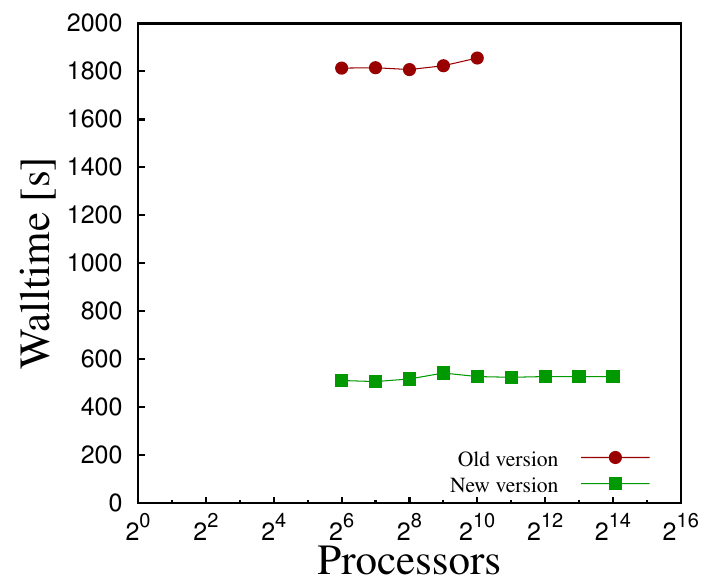}
		\end{subfigure}
		\hspace{0.01\textwidth}
		\begin{subfigure}[t]{0.48\textwidth}
			\includegraphics[width=\textwidth]{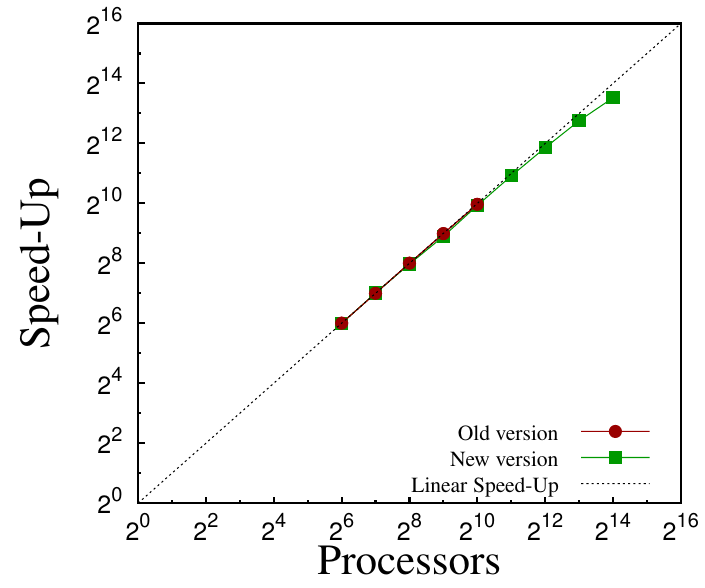}
		\end{subfigure}
\caption{Left: walltime for a single RK4 cycle in a cubic domain with a discretization of $2048^3$ grid points. Walltime $= t n$, where $t$ is the real time needed for the computation, and $n$ is the number of used processors. Right: Speed-up of the code. Speed-Up $= n_R t_R/t$, where $t_R$ and $n_R$ are reference quantities (in this case, $n_R = 64$).}
\label{fig:wall_time}
\end{figure}

%In order to make the simulations feasible (breaking down limitations due to computation time and memory required), the computational code uses a stencil parallelization to split the grid among a very large  amount of processors (according to the distributed memory paradigm), and communication are performed through the MPI libraries.

%Implementation and parallelization
{The code is based on TurIsMi, v1.4, of the Philofluid group (www.polito.it/philofluid), which was released under the terms of the GNU General Public License. A new version of the code has here been implemented using the Fortran 2018 standard. The new features allowed us to design the code as slightly object-oriented, thereby increasing the readability and efficiency of shared routines. Direct/inverse FFTs (Fast Fourier Transforms) are evaluated using FFTW (Fast Fourier Transforms of the West) open-source libraries (which support the shared memory paradigm).
Parallelization is performed with a hybrid (shared/distributed) memory paradigm. In particular, we have used a stencil parallelization (parallelization over two directions) to distribute the computational domain over a chosen number of processes (up to $N^2/2$ – theoretical value). This distribution was performed using the MPI 3.0 standard, which allows modern MPI libraries (such as OpenMPI and MPICH2) to be used.
In order to perform FFTs along a given direction, a process needs to know the values associated with all the wave-numbers in such a direction. For this purpose, matrix transpositions are mandatory to swap the distributed direction. During the inverse transform/transposition process, the domain is "expanded" by including the zero-padded anti-aliasing region (and viceversa, it is "contracted" during direct transforms). Using the expanded domain in a physical space only reduces the number of needed transforms. For simplicity, we considered a cubic domain, with $N^3$ in wavenumber space, and $M^3 = 27/8 N^3$ points in    physical space.
%Without the expansion/contraction process, the number of single FFTs required to perform a global transform would be equal to $3M^2 =27/4 N^2$.
Without the expansion/contraction process, the number of single FFTs required to perform a global transform would be equal to $N^2 + NM + M^2 =27/4N^2$, thereby a saving of 30\% of computational time was achieved.
The MPI 3.0 standard allows us to implement a global communication subroutine for direct/inverse domain transposition, and also for input/output routines. The shared part of the parallelization is managed by OpenMP in the rest of the code.
As a result of the optimization, the new version of the code is about 5 times faster, and has a near-linear speed-up, which allowed us to fully exploit the potential of massively parallelized supercomputers, see Fig. \ref{fig:wall_time}. The simulations were performed on the TGCC Curie supercomputer, within PRACE project n$^\circ$ RA07732011, for a total of 3 million cpu-hours.}

\section{Results}\label{resStrat}

In this section, we analyze the simulated fields by comparing the results obtained for the different stable and unstable stratification cases. We analyze the statistical behavior of the velocity and scalar fields in subsection A. The formation of kinetic energy sub-layers in the mixing region is discussed in subsection B.  {The effects related to the entrainment process are presented in subsection B.1, while the anisotropy, dissipation and small-scale effects are discussed in subsections B.2 and B.3, respectively}.

\begin{figure}[bht!]
	\centering
	\includegraphics[width=0.65\textwidth]{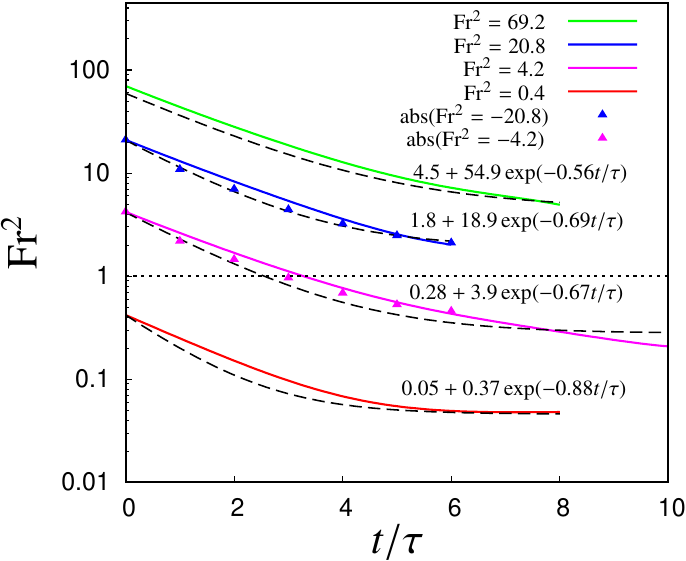}
	\caption{{Time evolution of the instantaneous Froude number \fr$(t)$ for the simulations with stable stratification. The evolution in unstable cases is very close to this one, except for the sign, which is negative. Dashed lines represent the fitting exponential laws of these temporal decays. For the readers' convenience, they have also been  gathered in Table III. The horizontal dotted line, \fr=1, indicates the moments in time when the buoyancy forces are of the same order as the inertial ones: the relative effects are found when \fr$(t)\approx 2\div3$. }}
\label{fig:froude}
\end{figure}

\begin{table}[bht!]
	\centering
	\begin{tabular}{|l|l|l|}
		\hline
		Froude number & Fitting parameters &  Asymptotic Standard Error\\
		\hline
		Fr$^2$ = 69.2 &  $b = 4.5, n = 54.9, u = 1.8$ & $\bigtriangleup_{b} = 5.5\%, \bigtriangleup_{n} = 1.7\%, \bigtriangleup_{u} = 2.8\%$ \\ \hline
		Fr$^2$ = 20.8 & $b = 1.8, n = 18.9, u = 1.44$ & $\bigtriangleup_{b} = 15.4\%, \bigtriangleup_{n} = 2.2\%, \bigtriangleup_{u} = 5.8\%$ \\ \hline
		Fr$^2$ = 4.2 & $b = 0.28, n = 3.9, u = 1.5$ & $\bigtriangleup_{b} = 10.5\%, \bigtriangleup_{n} = 1.7\%, \bigtriangleup_{u} = 3.9\%$ \\ \hline
		Fr$^2$ = 0.4 & $b = 0.05, n = 0.37, u = 1.13$ & $\bigtriangleup_{b} = 2.8\%, \bigtriangleup_{n} = 0.7\%, \bigtriangleup_{u} = 1.8\%$ \\ \hline
	\end{tabular}
	\caption{Exponential fits, $f(x) = b + n\exp^{-x/u}$, of the temporal decay of the Froude numbers shown in Fig. \ref{fig:froude}}
\end{table}

The temporal evolution of \fr\ is shown in Fig. \ref{fig:froude}. It is possible to see that the stratification is mainly enhanced because of the decay in the kinetic energy and the increase in the consequent integral scale.
\subsection{Spatial statistical properties}
\label{sec:stat}
The statistics are computed by averaging the variables in the planes $(x_1,x_2)$ normal to the mixing direction (with a sample of $2^{10}\times2^{10}$ data-points). We focus on the variation along the vertical (non-homogeneous) direction, $x_3$. We thus define the average operator $\langle\cdot\rangle(x_3)$ as the mean value inside a plane $(x_1,x_2)$ for given values of $x_3$:
$$
\langle {\mathbf{\cdot}} \rangle(x_3) = \frac{1}{2^{20}}\sum\limits_{i=1}^{2^{10}}\sum\limits_{j=1}^{2^{10}}\cdot\, (x_{1,i},x_{2,j},x_3).
$$ 

\noindent The second-order moment is represented by the variance in scalar fields $\theta$ and $\chi$ or by the turbulent kinetic energy of the velocity field, which is defined as 
$
\label{eq:kinEn}
E=\frac{1}{2}\left(\langle u^2_1\rangle + \langle u^2_2\rangle +\langle u^2_3\rangle\right)
$.
High-order moments are represented by skewness and kurtosis (third- and fourth-order moments normalized by means of  the variance), defined as 
%\begin{equation}
%\label{eq:highMom}
%S(\cdot)=\frac{\langle\cdot^3\rangle}{\langle\cdot^2\rangle^{1.5}}\ \ \ \ \ \ \ \ \ \ \ \ K(\cdot)=\frac{\langle\cdot^4\rangle}{\langle\cdot^2\rangle^{2}}.\ \ \ \ 
%\end{equation}
%\begin{equation}
%\label{eq:highMom}
$S(\cdot)={\langle\cdot^3\rangle}/{\langle\cdot^2\rangle^{1.5}}$ and  
$K(\cdot)={\langle\cdot^4\rangle}/{\langle\cdot^2\rangle^{2}}$, respectively.
It should be noted that the definition of skewness and kurtosis for passive scalar field $\chi$ differs slightly from the one given in the previous equation. Because of the proximity of the external regions, where the variance $\langle\chi^2\rangle$ vanishes, in order to prevent numerical problems, the actual definitions of skewness and kurtosis are modified as 
%\begin{equation}
%\label{eq:highMomChi}
%S(\chi)=\frac{\langle\chi^3\rangle}{\left(\langle\chi^2\rangle+0.01\langle\chi^2\rangle_M\right)^{1.5}}\ \ \ \ \ \ \ \ \ \ \ \ K(\chi)=\frac{\langle\chi^4\rangle}{\left(\langle\chi^2\rangle+0.01\langle\chi^2\rangle_M\right)^{2}}\ \ \ \ 
%\end{equation}
$
S(\chi)={\langle\chi^3\rangle}/{\left(\langle\chi^2\rangle+0.005\langle\chi^2\rangle_\mathrm{max}\right)^{1.5}}$ and $K(\chi)={\langle\chi^4\rangle}/{\left(\langle\chi^2\rangle+0.01\langle\chi^2\rangle_\mathrm{max}\right)^{2}}$, where $\langle\chi^2\rangle_\mathrm{max}$ indicates the maximum variance value along direction $x_3$.

\begin{figure}[bht!]
	\begin{minipage}{.5\textwidth}
		\centering
		\includegraphics[width=\textwidth]{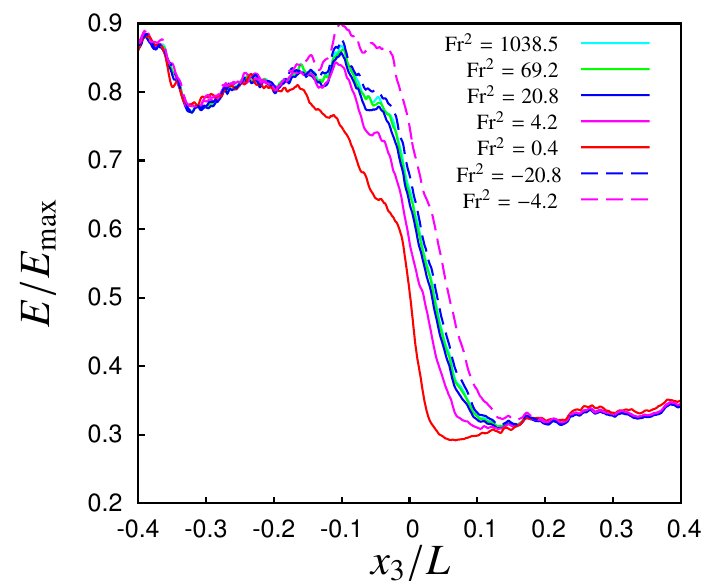}
		\hspace*{1cm}
		(\textit{a})\ Kinetic energy, $t/\tau = 3$
	\end{minipage}\hfill
	\begin{minipage}{.5\textwidth}
		\centering
		\includegraphics[width=\textwidth]{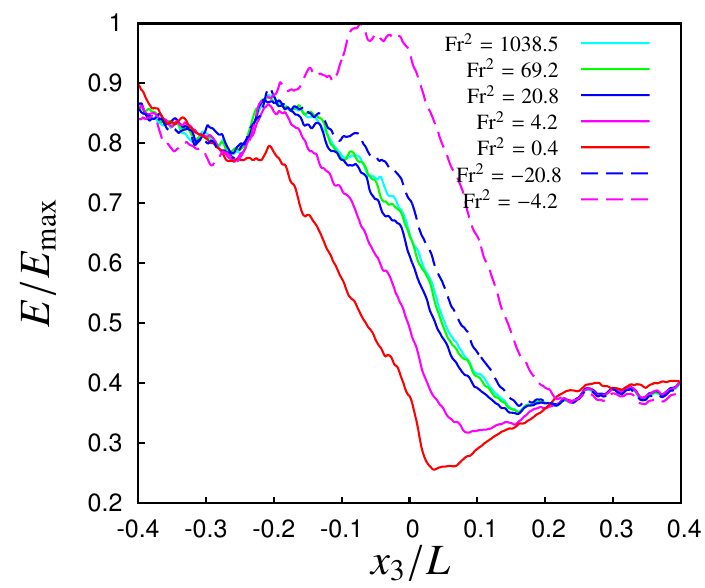}
		\hspace*{1cm}
		(\textit{b})\ Kinetic energy, $t/\tau = 6$
	\end{minipage}	
	\caption{Turbulent energy along vertical direction $x_3$, computed from the velocity variance in the horizontal planes, $x_1-x_2$.  The data are taken from simulations with different levels of stratification, which are represented by the initial reference squared Froude number \fr.}
	\label{fig:enerVel}
\end{figure}

%%%%%%%%%%%%%%
%  FIGURE 3  %
%%%%%%%%%%%%%%
As a result of the evolution of the ratio between the buoyancy force and the other dynamical  effects (advection and diffusion) and by looking at the statistical behavior of the turbulent kinetic energy shown in Fig. \ref{fig:enerVel}, the evolution of the system can be split into two main stages. As long as the ratio remains small, no significant differences emerge with respect to a non-stratified case. However, as the stratification perturbation becomes more important, buoyancy effects prevail, and  differences are present from both the quantitative and qualitative points of view.
The effects of different stratification levels are clearly visible on the turbulent kinetic energy shown in Fig. \ref{fig:enerVel}, where two different instants are compared,  $t/\tau=3$ in Fig. \ref{fig:enerVel} (a) and $t/\tau=6$ in Fig. \ref{fig:enerVel} (b), where $\tau$ is the initial eddy turnover time.
When the buoyancy term becomes comparable with the other forces, %depends  on the initial stratification intensity. When this condition is reached, initially there is 
a slight downward displacement of the energy gradient location takes place. %see curves for \fr=4.2 and \fr=0.4 at $t/\tau=3$ (Fig. 5a) .%and for \fr=20.8 at . 
Subsequently, %see $t/\tau=6$ (Fig. 5b), 
the onset of a sub-layer, characterized by a widening of the pit of kinetic energy in time, can be observed, see also Section \ref{sec:energyPit} and Fig. \ref{fig:thicknesses} (panels c,d,e).
The presence of such a sublayer changes the system dynamics, because two interfaces are produced in this situation. The first - which would also be present in the absence of stratification - separates the high turbulent energy region from the pit. The second one - which would not be present without stratification - separates the low turbulent energy region from the center of the mixing layer. Therefore, a strong stable stratification induces a kind of physical separation between the regions below and above the mixing layer, thus decreasing their interaction to a great extent. On the other hand, an increment of the kinetic energy inside the mixing region, a sort of peaky sublayer, can be observed in unstable cases. Again in this case, we observe the formation of a secondary energy gradient, but its location is reversed with respect to the stable case. At this point, the secondary gradient separates the peak from the high-energy region where the vapor cloud is located. In fact, the peak is shifted toward the high energy region (while the pit is closer to the low energy one). The principal gradient is now pushed upward (positive $x_3$), see also panel (e) in Fig. \ref{fig:thicknesses}.  % and separates the peak from the low energy clear-air region.

%%%%%%%%%%%%%%%%%%%%%%%%%%
% FIGURE 4 (vel ske/kur) %
%%%%%%%%%%%%%%%%%%%%%%%%%%

\begin{figure}[bht!]
	\begin{minipage}{.5\textwidth}
		\centering
		\includegraphics[width=\textwidth]{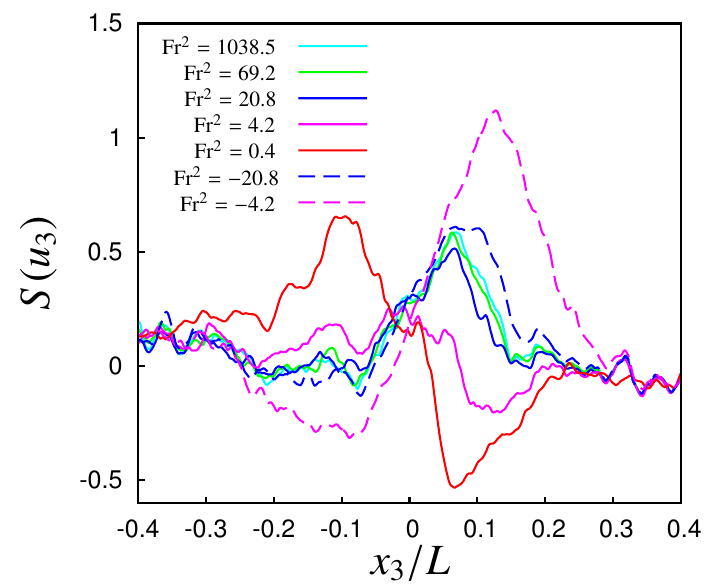}
		\hspace*{1cm}
		(\textit{a})\ Vertical velocity skewness, $t/\tau = 6$
	\end{minipage}\hfill
	\begin{minipage}{.5\textwidth}
		\centering
		\includegraphics[width=\textwidth]{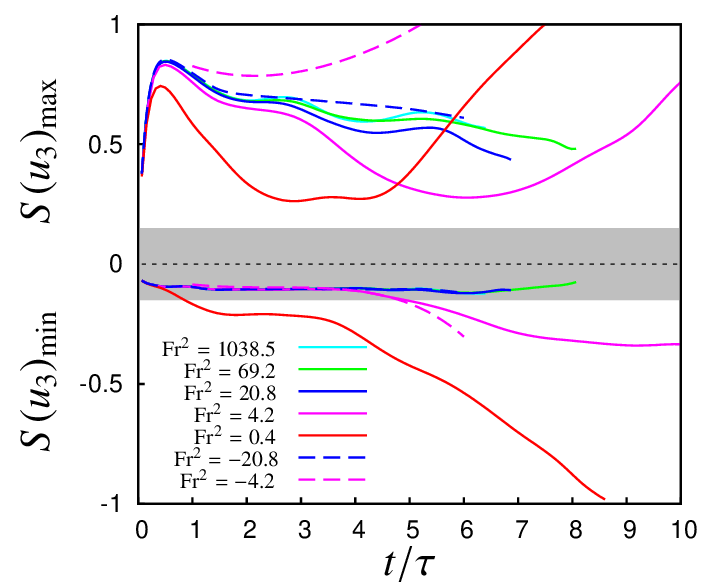}
		\hspace*{1cm}
		(\textit{b})\ Peak values of the vertical velocity skewness over time
	\end{minipage}\hfill
	\begin{minipage}{.5\textwidth}
		\centering
		\includegraphics[width=\textwidth]{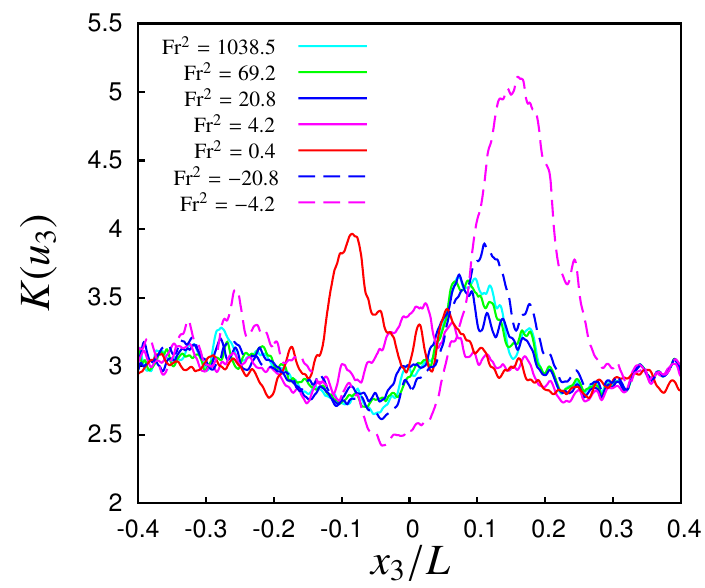}
		\hspace*{1cm}
		(\textit{c})\ Vertical velocity kurtosis	, $t/\tau = 6$
	\end{minipage}\hfill
	\begin{minipage}{.5\textwidth}
		\centering
		\includegraphics[width=\textwidth]{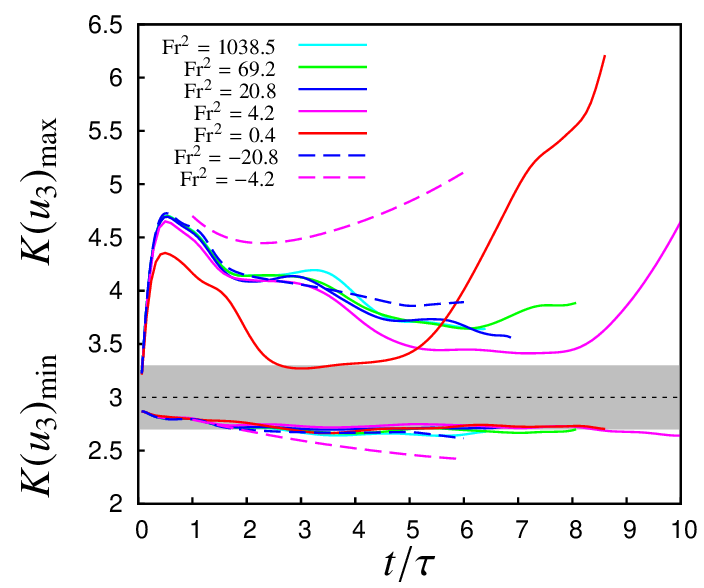}
		\hspace*{1cm}
		(\textit{d})\ Peak values of the vertical velocity kurtosis over time
	\end{minipage}\hfill
	\caption{Vertical velocity skewness (panels \textit{a-b}) and kurtosis (panels \textit{c-d}) along vertical direction $x_3$, computed from vertical velocity central moments in the horizontal planes $(x_1-x_2)$.  The data are taken after 6 $\tau$ (panels \textit{a-c}) and over the temporal evolution  (panels {b-d}). Simulations with a different stratification are represented by the square Froude number, \fr. The gray band in panels b and d represents the intermittency range measured outside the mixing layer.}
	\label{fig:skewVel}
\end{figure}

The parts of the flow where the primary energy gradient acts and the secondary one (when present), behave intermittently. Fig. \ref{fig:skewVel} shows  the   skewness and kurtosis of the vertical velocity fluctuations after 6 time scales (panels \textit{a} and \textit{c}, respectively), and the time evolution of their maximum and minimum values (panels \textit{b} and \textit{d}).
A reduction in the maximum values, which decay much faster than the non-stratified or weakly stratified cases, can be observed for the stably stratified cases, beyond $t/\tau \sim 1$. Such a fast decay during pit formation leads to a low intermittency, which is characterized by values as low as those observed outside the mixing anisotropic  region (the "normal" range is represented by a gray band in panels \textit{b} and \textit{d} in Fig.\ref{fig:skewVel}). $S$ and $K$ then grow quickly in time, reaching higher values than the unstratified case. The  intermittency decay in the unstable stratification case is immediately damped and a growth of $S$ and $K$ is observed for the \fr=-4.2 case beyond 3 time scales. The final configuration at the end of the numerical simulation seems to be more  intermittent in both the stable and unstable cases, with values that can become even  $100\%$ larger than in the unstratified case. 

%%%%%%%%%%%
% SCALARS %
%%%%%%%%%%%

\begin{figure}[bht!]
	\centering
	\begin{subfigure}[t]{0.39\textwidth}
		\includegraphics[width=\textwidth]{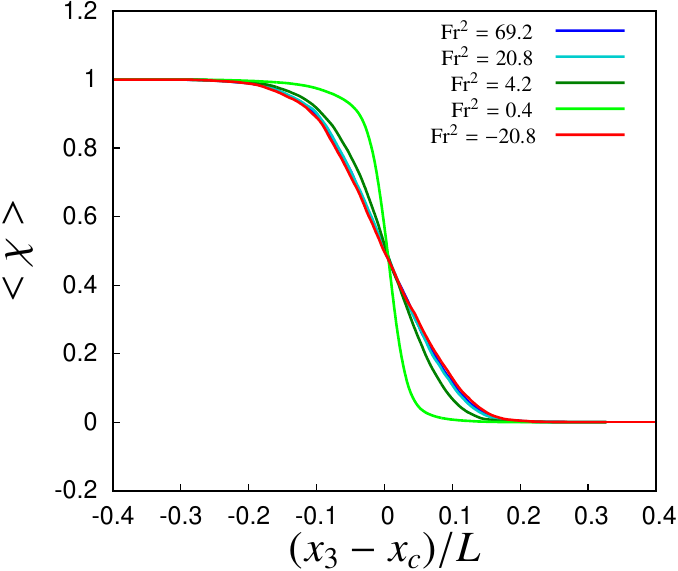}
	\end{subfigure}
	\hspace{0.01\textwidth}
	\begin{subfigure}[t]{0.39\textwidth}
		\includegraphics[width=\textwidth]{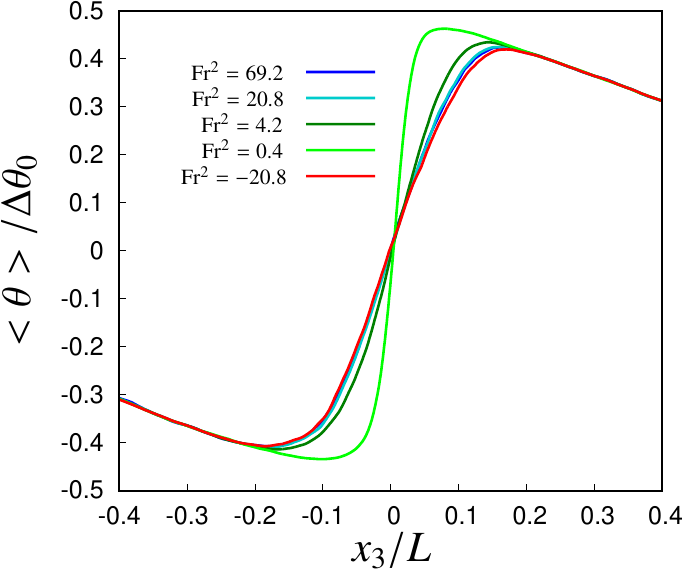}
	\end{subfigure}
	
	\vspace{0.3cm}
	
	\begin{subfigure}[t]{0.39\textwidth}
		\includegraphics[width=\textwidth]{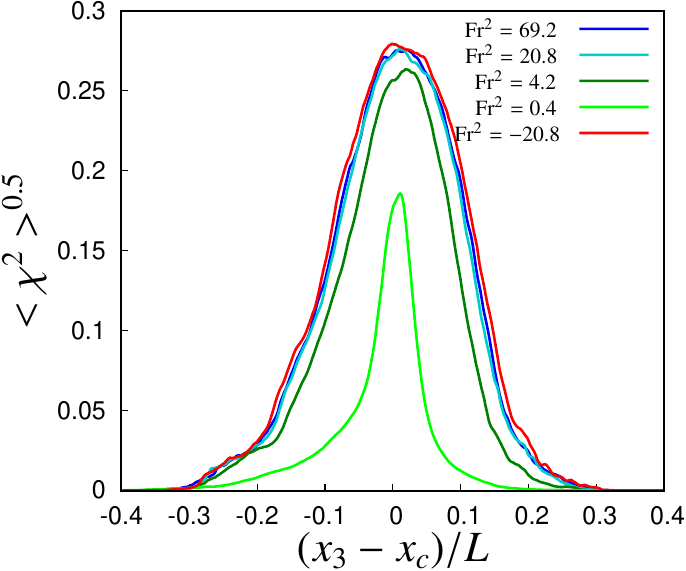}
	\end{subfigure}
	\hspace{0.01\textwidth}
	\begin{subfigure}[t]{0.39\textwidth}
		\includegraphics[width=\textwidth]{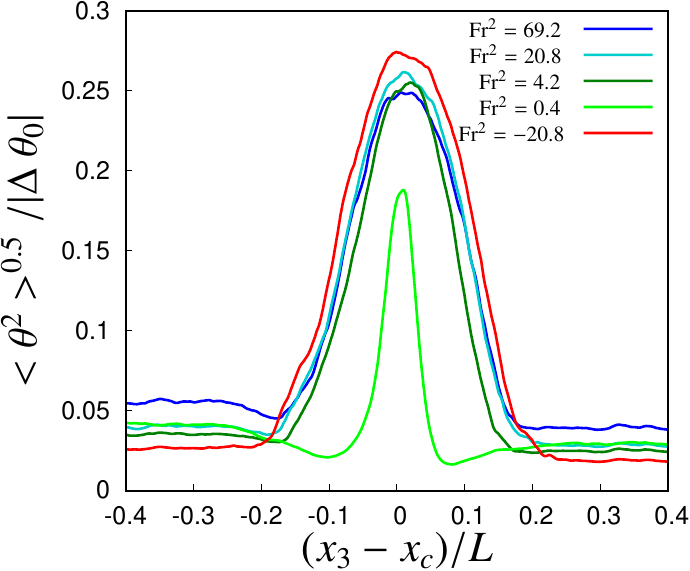}
	\end{subfigure}
	
	\vspace{0.3cm}
	
	\begin{subfigure}[t]{0.39\textwidth}
		\includegraphics[width=\textwidth]{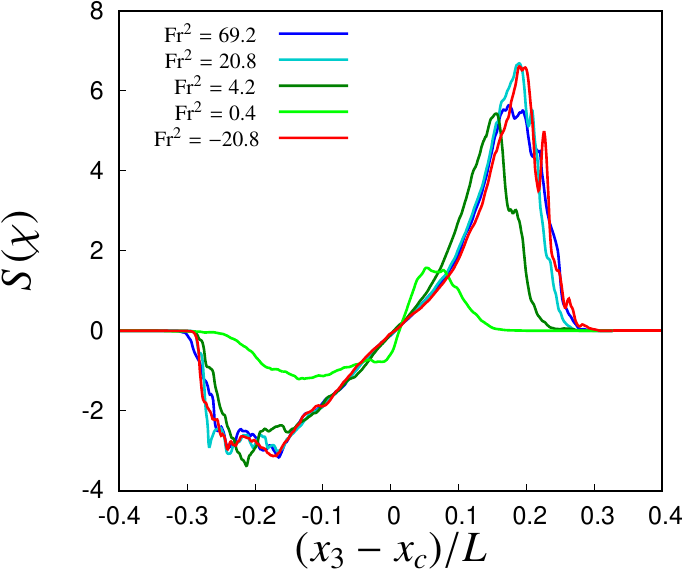}
	\end{subfigure}
	\hspace{0.01\textwidth}
	\begin{subfigure}[t]{0.39\textwidth}
		\includegraphics[width=\textwidth]{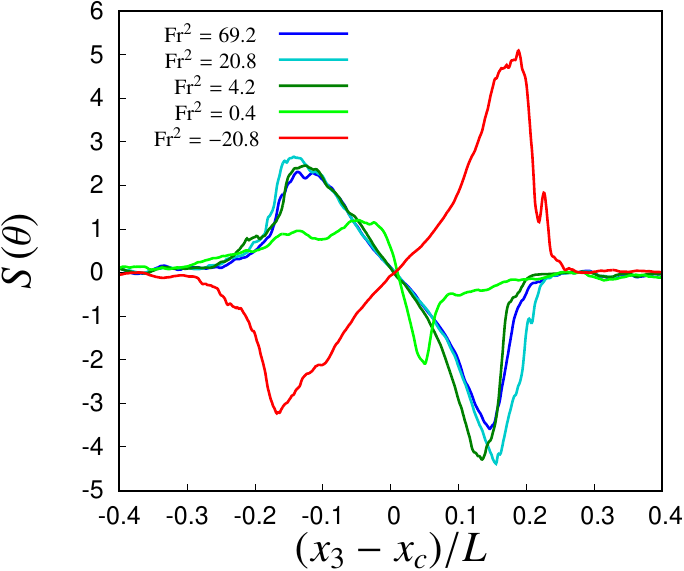}
	\end{subfigure}
	
	\vspace{0.3cm}
	
	\begin{subfigure}[t]{0.39\textwidth}
		\includegraphics[width=\textwidth]{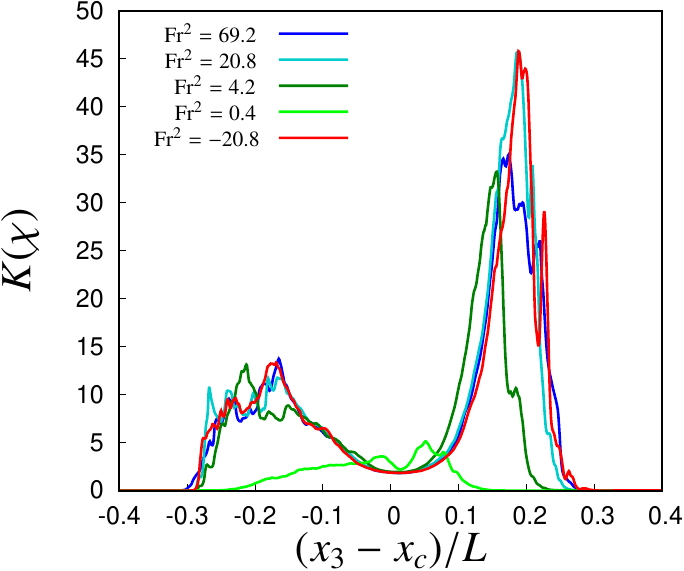}
	\end{subfigure}
	\hspace{0.01\textwidth}
	\begin{subfigure}[t]{0.39\textwidth}
		\includegraphics[width=\textwidth]{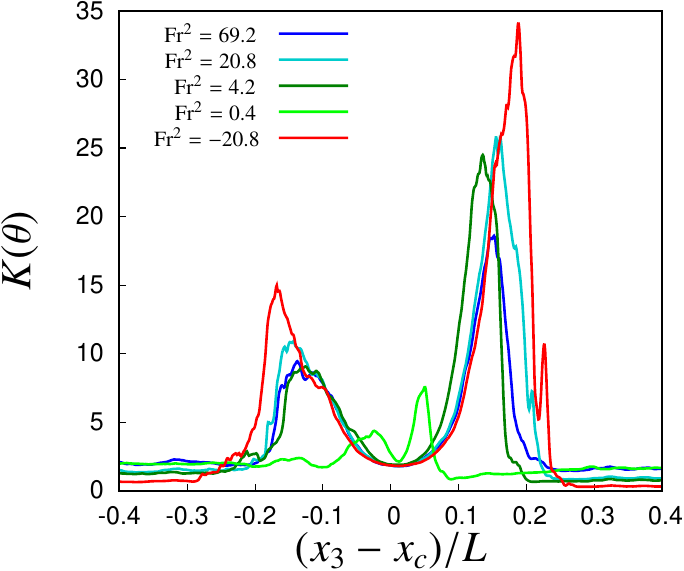}
	\end{subfigure}
	\caption{{Comparison of the passive (left) and active (right) scalar statistics, $t/\tau = 6$.}}
	\label{fig:stat-passive-activeScal} 
\end{figure}

The statistical properties of temperature fluctuation $\theta'$, of the active scalar, and of the vapor passive scalar concentration, $\chi$, are analogous. In fact, the non-differential term $u_3 G_0$ in energy equation \ref{eq:ene} does not exert an effect that is comparable with that of the buoyancy term inside the momentum equation. The latter has a vectorial nature and efficiently receives and transposes the gravitational effect to the velocity field. Our simulations show that the transport of temperature is comparable with the transport produced by advective-diffusive equation \ref{eq:scal} in the vapor field, that is, a simple passive scalar field, see Fig. \ref{fig:stat-passive-activeScal}, where the first four statistical moments across the interface are presented at $t/\tau=6$. %(with the Schmidt number instead of the Prandtl number).
The effects on the scalar fields are milder than the ones observed on the velocity. The width of the region with non-zero variance depends on the stratification level and becomes thinner for stable cases. Substantial variations can be observed for the case of very strong stratification, for example when \fr=0.4. Scalar fluctuations are damped in stable cases, and slightly enhanced in the presence of unstable stratification. %, as can be observed in Fig. \ref{fig:statScal}(\textit{a}).  % - which indicates a smaller mixing region. 
%It can be also observed that the shape of mean variance along $x_3$ is clearly non-symmetrical with respect to the mixing center.
The shrinking of the mixing layer becomes remarkable after the onset of the pit of energy and is linked to the reduction in entrainment,  see subsection B.1 and, %Fig. \ref{fig:statScal}(\textit{b}) where the evolution of scalar variance peak value is shown. 
for a complete overview, see Gallana's   PhD thesis \cite{gallana16}.

As far as the high-order moments are concerned, the scalar fields initially follow the same trend  as the velocity fluctuations, with a reduction in $S$ and $K$ when the stratification is stable and a growth when it is unstable, see the panels in the second, third and fourth rows in Fig. \ref{fig:stat-passive-activeScal}.  A large difference can be observed, after a few time scales, in the stable stratification case. Here, the onset of the energy pit blocks the mixing process, and the values of the high-order statistics tend to remain almost constant. %, as shown in Fig.  \ref{fig:statScal}, panels (\textit{d-f}). As for the variance of temperature fluctuations, substantial variations on $ S $ and $ K $ are visible for \fr = 0.4 only, while the other cases show minor variations in behavior.

\begin{figure}[bht!]
		\centering
		\begin{subfigure}[t]{0.4\textwidth}
			\includegraphics[width=\textwidth]{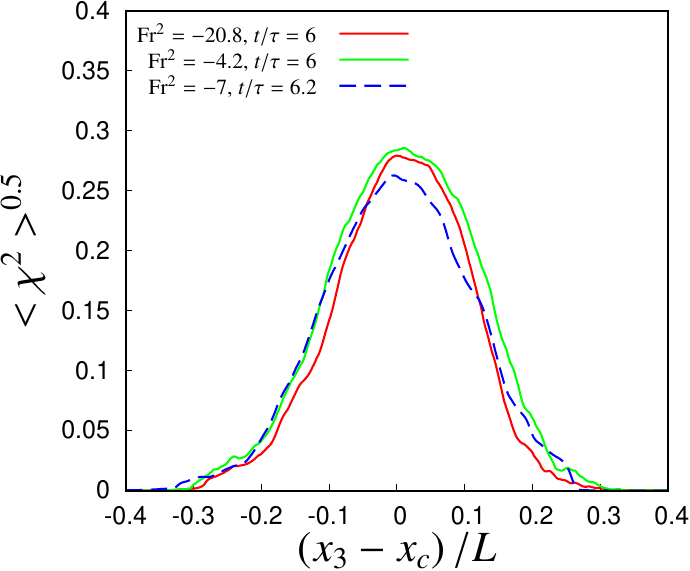}
		\end{subfigure}
		\hspace{0.01\textwidth}
		\begin{subfigure}[t]{0.4\textwidth}
			\includegraphics[width=\textwidth]{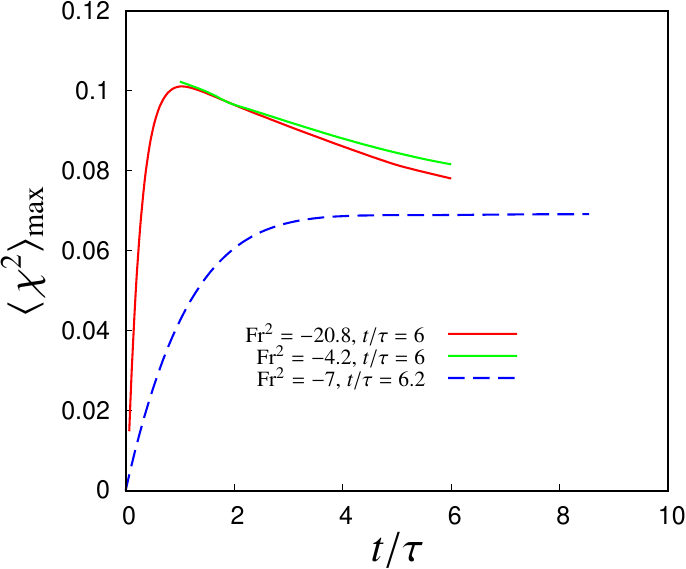}
		\end{subfigure}
		
		\vspace{0.3cm}
		
		\begin{subfigure}[t]{0.4\textwidth}
			\includegraphics[width=\textwidth]{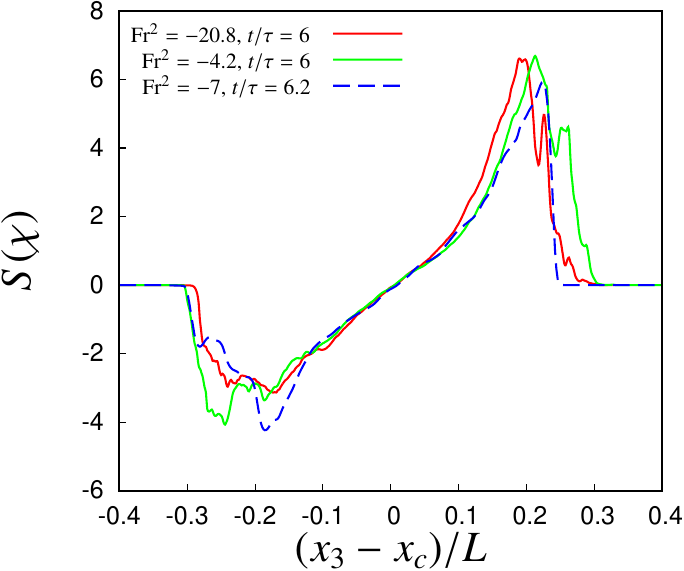}
		\end{subfigure}
		\hspace{0.01\textwidth}
		\begin{subfigure}[t]{0.4\textwidth}
			\includegraphics[width=\textwidth]{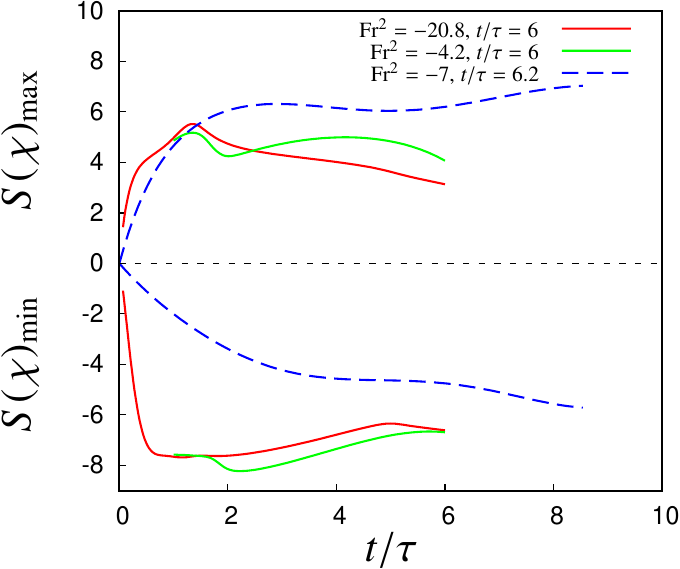}
		\end{subfigure}
		
		\vspace{0.3cm}
	
		\begin{subfigure}[t]{0.4\textwidth}
			\includegraphics[width=\textwidth]{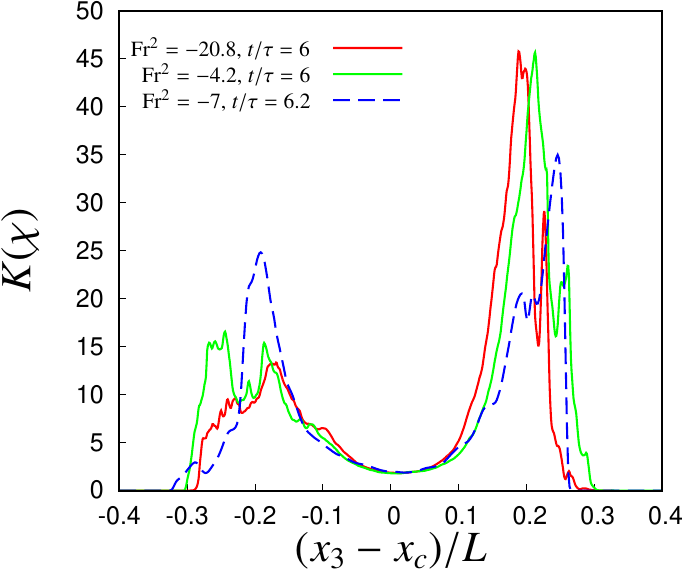}
		\end{subfigure}
		\hspace{0.01\textwidth}
		\begin{subfigure}[t]{0.4\textwidth}
			\includegraphics[width=\textwidth]{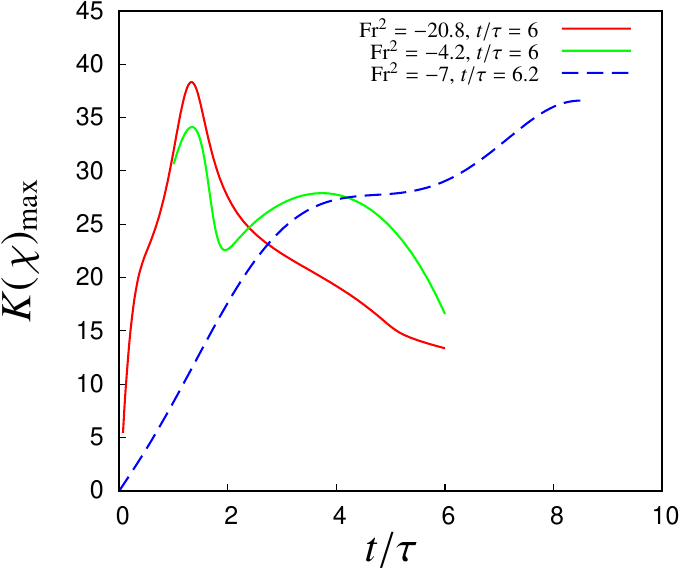}
		\end{subfigure}
		\caption{ {Comparison of vapor moment statistics for simulations with droplets (dashed) and without (solid, same data as shown in the left column of Fig. \ref{fig:stat-passive-activeScal}).}}
		\label{fig:statScal-drops}
	\end{figure}

{It is also interesting to note that the morphology of the spatial distribution of the vapor statistics, the passive scalar, is not affected to any great extent by the presence of a population of either monodisperse or polydisperse water drops. Indeed, if a comparison is made between our simulations containing the aqueous phase, which is equivalent in quantity to what is present inside warm clouds (LWC, Liquid Water Content, equal to $0.8 gr/m^3$), see Golshan et al. 2021 \cite{golshan2021} and the work of Fossa' et al. 2022 \cite{fossa2022}, which was carried out under almost the same Froude numbers, it can be seen that only the temporal evolution of the maximum and minimum peaks of the vapor statistical distributions are in fact affected, albeit only slightly, by the presence of drops, and by the related phenomenology  of  evaporation-condensation and collision-coalescence, see Fig.\ref{fig:statScal-drops}. Moreover, there is a variation of the maximum values of the Kurtosis function, which does not settle, in the long term, on the same asymptotic values, see the bottom right panel in Fig.\ref{fig:statScal-drops}.
Furthermore, it can be observed that the thinning of the mean temperature profile when the stratification is increased, which thins by about four times as the stratification increases from neutral to Fr$=0.4$, i.e. 5000 G$_0$, is the same as that measured by Jayesh and Warhaft\cite{jw94}, see Fig. 7, and in particular the curve where the mean temperature profile half width is  normalized by the integral lengthscale of the large-scale turbulence on the
lower side of their mixing layer. }

\begin{figure}[bht!]
	\centering
	\includegraphics[width=0.8\textwidth]{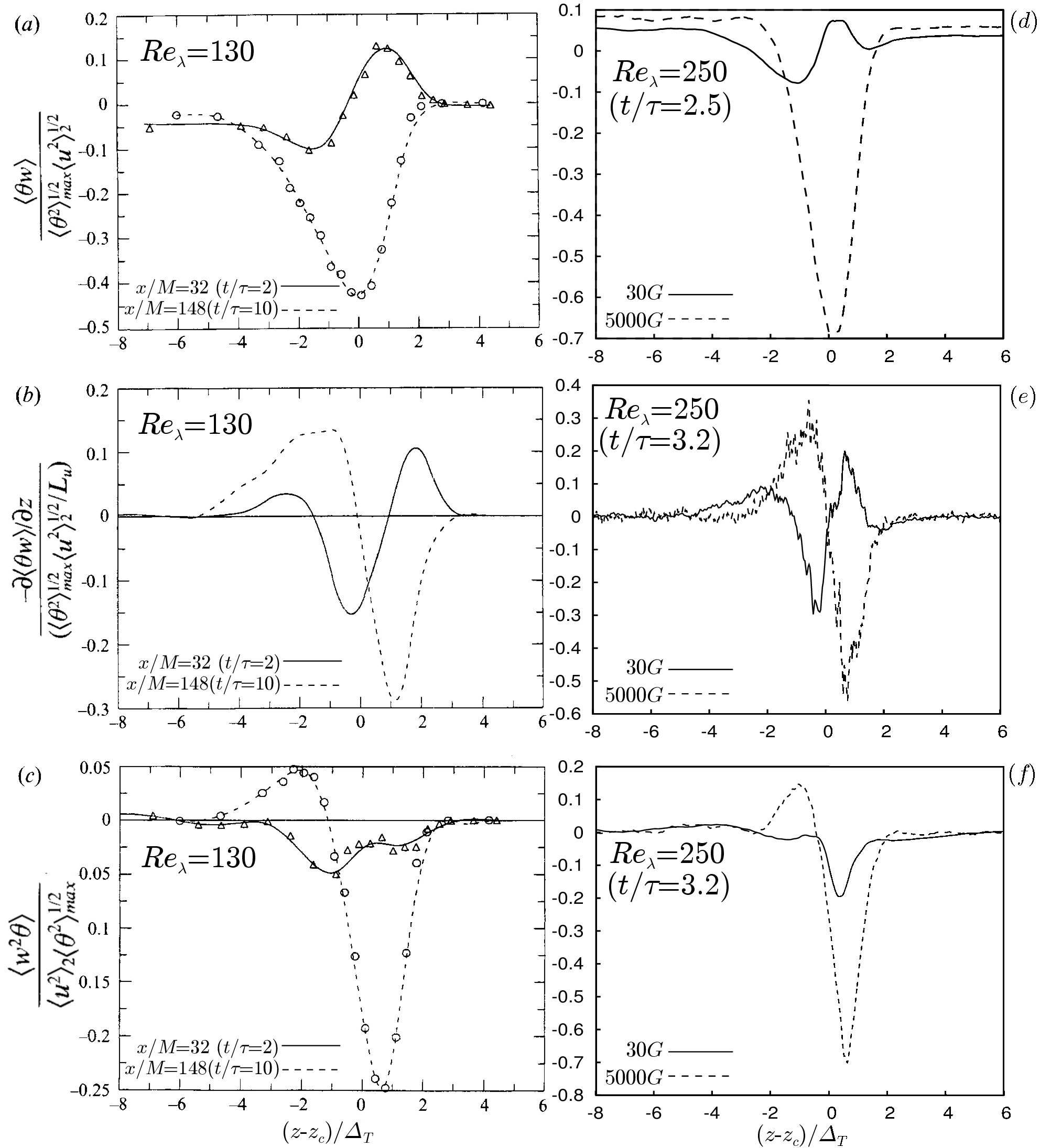}
	\caption{{Comparison of the flow statistics between the Jayesh and Warhaft laboratory experiment (J. Fluid Mech. 277 (1994), p. 29) (left column, spatial evolution) and the present numerical experiment (right column, temporal evolution). In their experiment, Jayesh and Warhaf considered a turbulent mixing between two regions with different kinetic energies and temperatures (see Figure 2). The velocity fluctuations were generated by forcing a flow into grids of different mesh sizes. Jayesh and Warhaft's data refer to $Ri = 0.8 (x/M=32$, dashed line) and $Ri = 63 (x/M = 148$, solid line). By using a Taylor transformation, it is possible to see that $x/M=32$ corresponds to a 2 time scale long temporal evolution, while x/M=148 corresponds to a 10 time scale long temporal evolution. The flows simulated in this work refer to $Ri = 0.11$ (30G case, where \fr = 69.2, dashed line) and to $Ri = 18.2$ (5000G case, where \fr = 0.4, solid line). In panel (d), $t/\tau = 2.5$, while in panels (e) and (f), $t/\tau = 3.2$. Panels (a) and (d): temperature flow. Panels (b) and (e): derivative normal to the mixing of the temperature flux. Panels (c) and (f): the temperature fluctuation flux (correlation between the second-order moment of the velocity fluctuation across the layer and the temperature fluctuation).}}
	\label{fig:comparisonJayesh-Warhaft}
\end{figure}

{As validation of our simulations, we present a comparison with the results of a similar  laboratory study carried out by Jayesh and Warhaft (JW in the following) in 1994 at Cornell University \cite{jw94}. In their experiment, a stably stratified interface, with strong turbulence below and quiescent air above, was studied in a wind tunnel with the aim of simulating the conditions at the inversion cap at the top of the atmospheric boundary layer. 
Thus, this system is the same as the one in our study as regards the transport of momentum, turbulent energy and temperature, although the transport of the passive scalar is missing. They generated the interfacial layer by means of a composite grid, with a small mesh size above and a large one below, see \cite{jw94}. A remarkable similarity can be observed in Fig. \ref{fig:comparisonJayesh-Warhaft} in the distributions across the mixing layer of the fluctuations of the temperature flux, of its spatial derivative and of the covariance between $ <u_3 ^ 2 \theta'> $ (note that in JW $x_3$ is represented by $z$ and $u_3$ by $w$). Only the extreme values are different as different Reynolds$_ {\lambda}$ values are considered, that is, equal to 130 in the JW's work laboratory and 250 in our numerical simulations. The reverse sign of the flux of the temperature inside the weak turbulence region, which corresponds to a counter-gradient heat flux (see also Riley, Metcalfe and Weissmann 1981\cite{riley81}, and Yoon and Warhaft 1990\cite{yoon}) should be noted in particular. 
The correspondence of the trends across the layer between our numerical experiment and the laboratory ones of JW extends to the kinetic energy flow, see subsection B.1 and Figure 17 in JW and our Figure 14. A dynamic aspect, which accompanies the formation of the kinetic energy pit and the blockage of the mixing layer growth, can be observed. }

\begin{figure}[bht!]
	\centering
	\begin{subfigure}[t]{0.4\textwidth}
		\includegraphics[width=\textwidth]{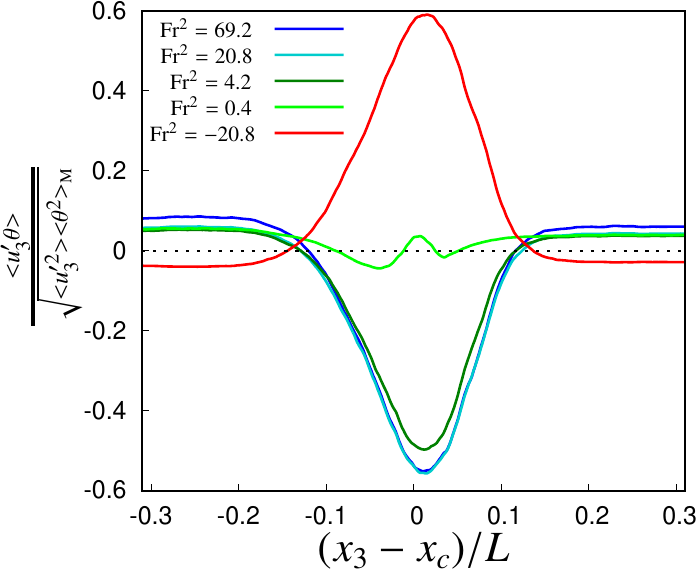}
		\caption{temperature flux}
	\end{subfigure}
	\hspace{0.01\textwidth}
	\begin{subfigure}[t]{0.4\textwidth}
		\includegraphics[width=\textwidth]{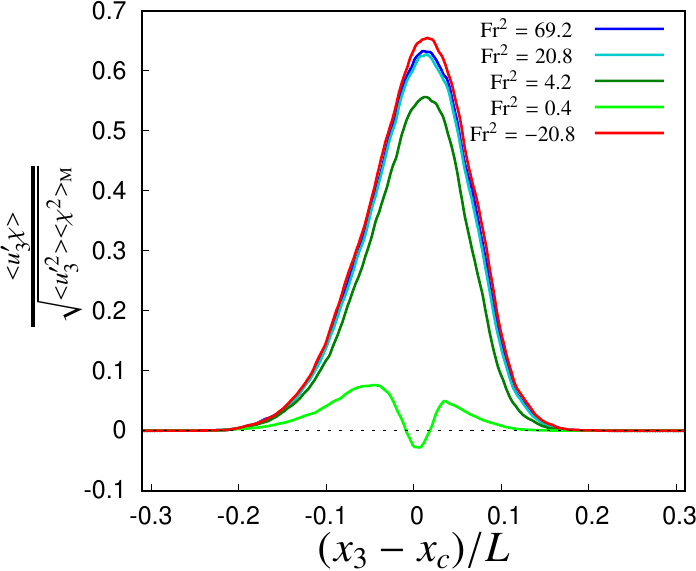}
		\caption{vapor flux}
	\end{subfigure}
	
	\vspace{0.3cm}
	
	\begin{subfigure}[t]{0.4\textwidth}
		\includegraphics[width=\textwidth]{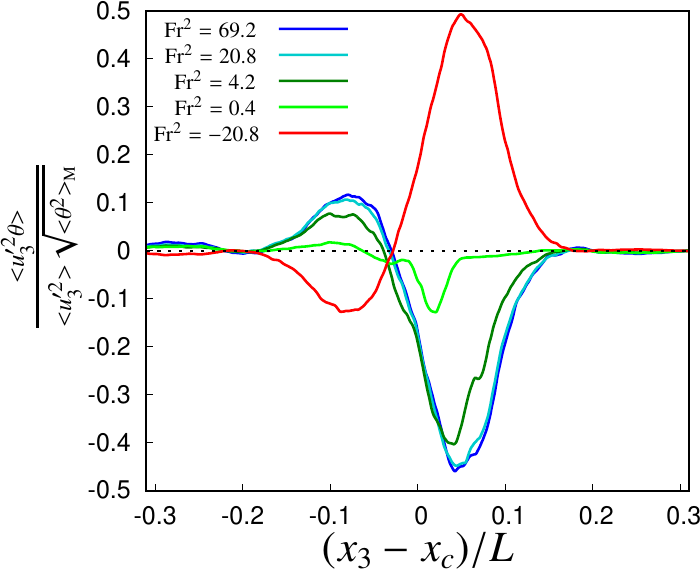}
		\caption{temperature flux transport}
	\end{subfigure}
	\hspace{0.01\textwidth}
	\begin{subfigure}[t]{0.4\textwidth}
		\includegraphics[width=\textwidth]{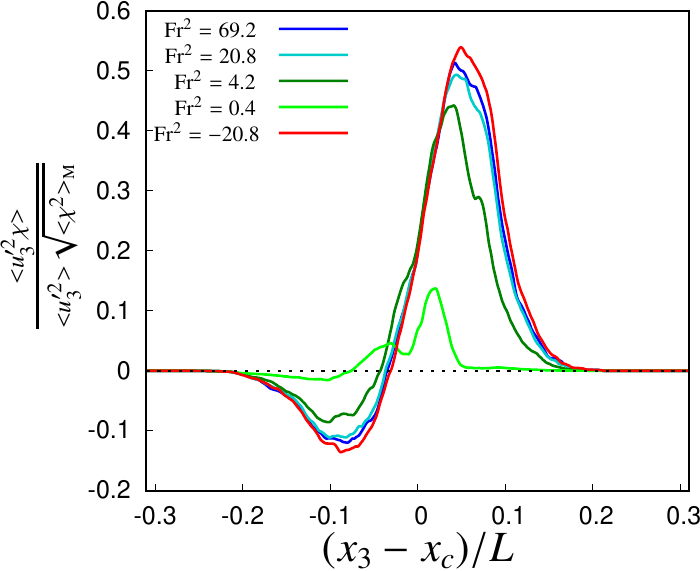}
		\caption{vapor flux transport}
	\end{subfigure}

	\vspace{0.3cm}
	
	\begin{subfigure}[t]{0.4\textwidth}
		\includegraphics[width=\textwidth]{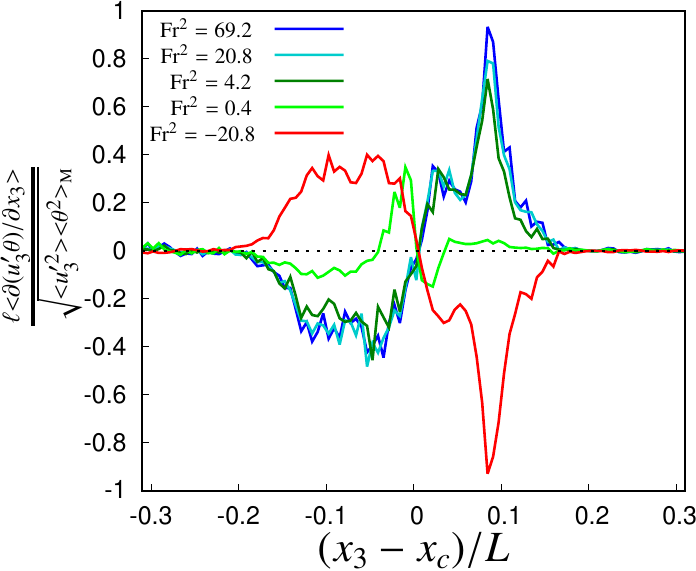}
		\caption{temperature flux derivative across the interface}
	\end{subfigure}
	\hspace{0.01\textwidth}
	\begin{subfigure}[t]{0.4\textwidth}
		\includegraphics[width=\textwidth]{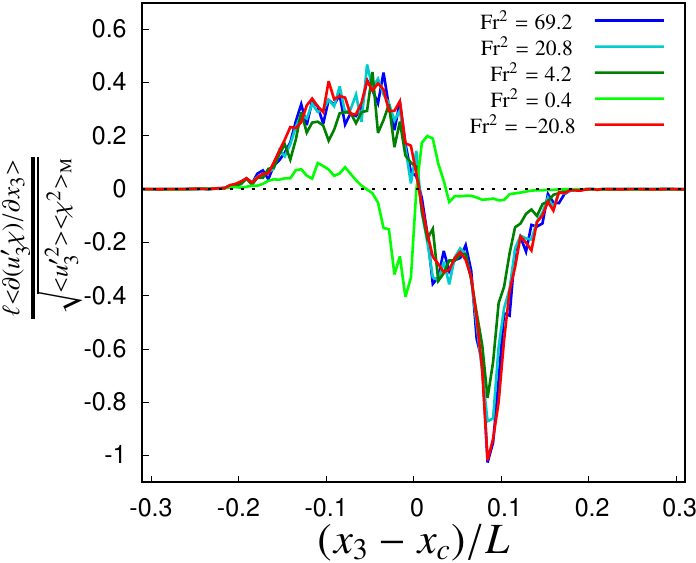}
		\caption{vapor flux derivative across the interface}
	\end{subfigure}

	\caption{{Comparison of the normalized heat (active scalar) and vapor (passive scalar) flux profiles across the interface, their vertical derivatives, and their flux for different levels of stratification, $t/\tau=3$.}}
	\label{fig:fluxes-active-passive-scalars}
\end{figure}

{To complete this section, we report a comparison between the active and passive scalars studied here in Fig. \ref{fig:fluxes-active-passive-scalars}, for the same distributions shown in Fig. \ref{fig:comparisonJayesh-Warhaft}. Once again, a remarkable similarity can be observed in the behavior of the two scalars, thus demonstrating the clear dominance of convective transport on the scalar, which, in principle, should be of the active type.}

%%%%%%%%%

%%%%%%%%%

%%%%%%%%%%%

%%%%%%%%%%%

\subsection{Stratified shearless turbulent mixing  and the formation of energy pit/peak sublayers}
\label{sec:energyPit}

\begin{figure}[bht!]
    \centering
	\begin{subfigure}[t]{0.32\textwidth}
		\includegraphics[width=\linewidth]{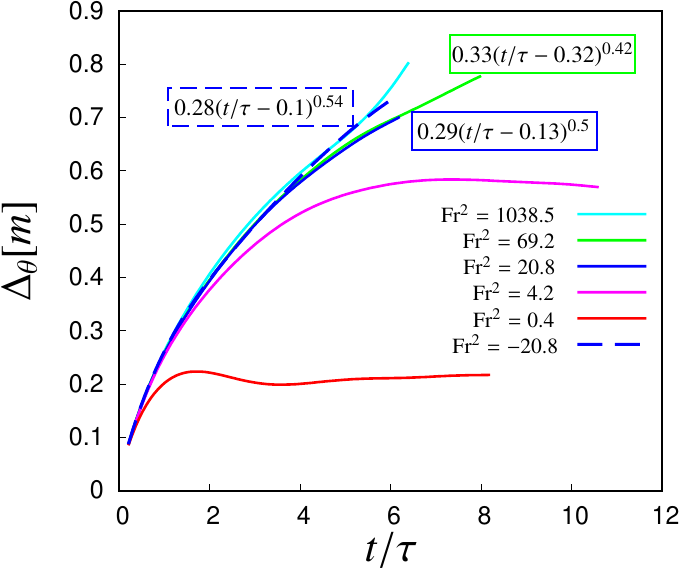}
		\caption{}
	\end{subfigure}
	\hfill
	\begin{subfigure}[t]{0.32\textwidth}
		\includegraphics[width=\linewidth]{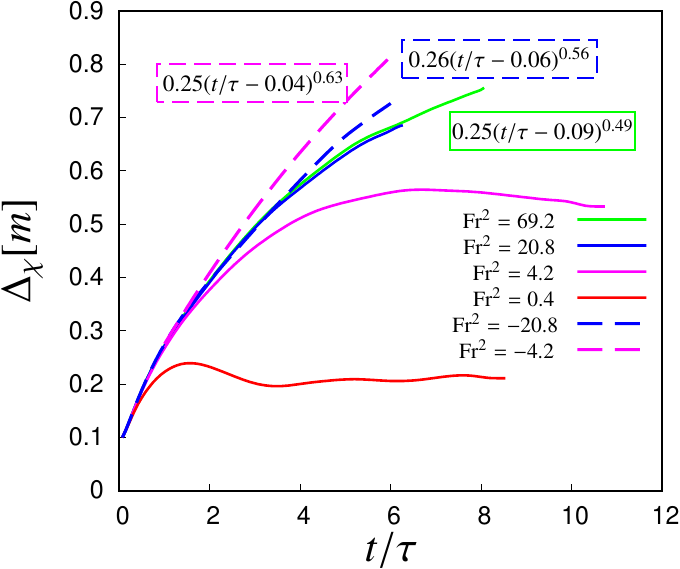}
		\caption{}
	\end{subfigure}
	\hfill
	\begin{subfigure}[t]{0.3\textwidth}
		\begin{tcolorbox}[colback=blue!5,colframe=olive,width=\linewidth,title={\small Thickness of the mixing layer}]
		\begin{itemize}[leftmargin=*,before=\setlength{\baselineskip}{5pt},itemsep=-3pt]
		{\tiny 
		\item $\bigtriangleup_{\chi,\theta} = x_{3,1} - x_{3,2}$
		\item $x_{3,1}$ is the water vapor front (where $\left\langle\chi,\theta\right\rangle = 0.25$)
		\item $x_{3,2}$ is the clear air front (where $\left\langle\chi,\theta\right\rangle = 0.75$)
		\item {\color{OliveGreen} Thickening stops at the onset of the pit, while mild transient growth is observed when stratification is marginal}
		\item {\color{red} Overgrowth is observed for unstable cases}
		}
		\end{itemize}
		\end{tcolorbox}
		\caption*{}
	\end{subfigure}
	\normalsize
	\vspace{0.3cm}
	\begin{subfigure}[t]{0.32\textwidth}
		\includegraphics[width=\textwidth]{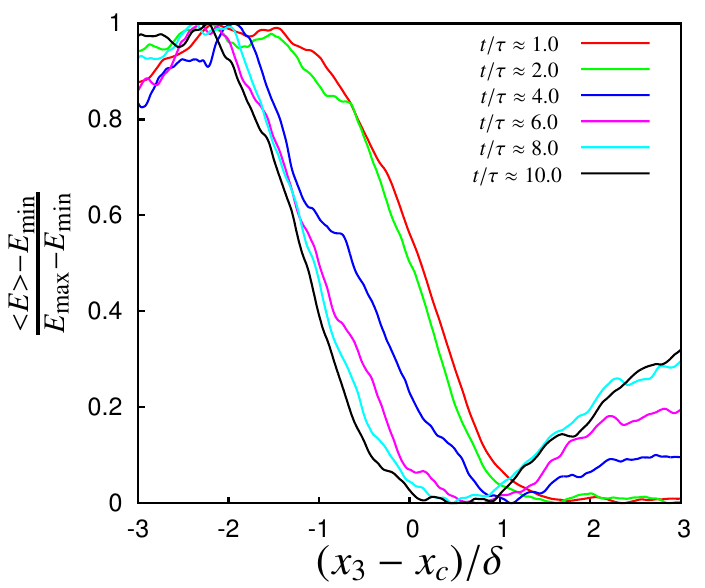}
		\caption{}
	\end{subfigure}
	\hfill
	\begin{subfigure}[t]{0.32\textwidth}
		\includegraphics[width=\textwidth]{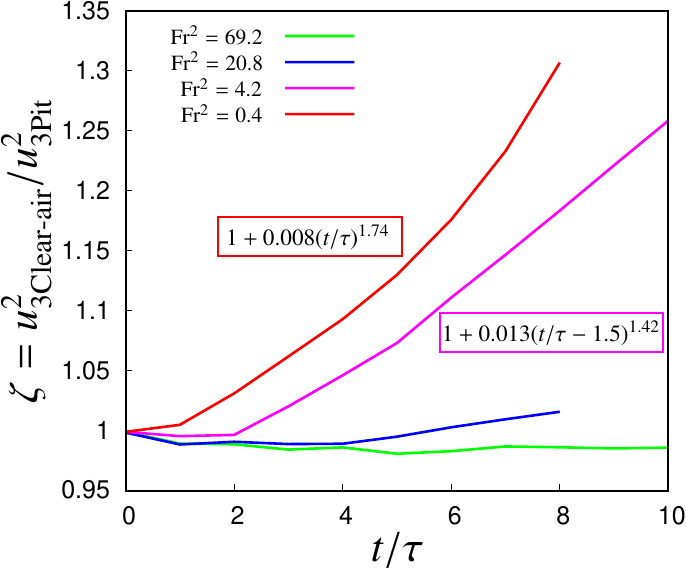}
		\caption{}
	\end{subfigure}
	\hfill
	\begin{subfigure}[t]{0.32\textwidth}
		\includegraphics[width=\textwidth]{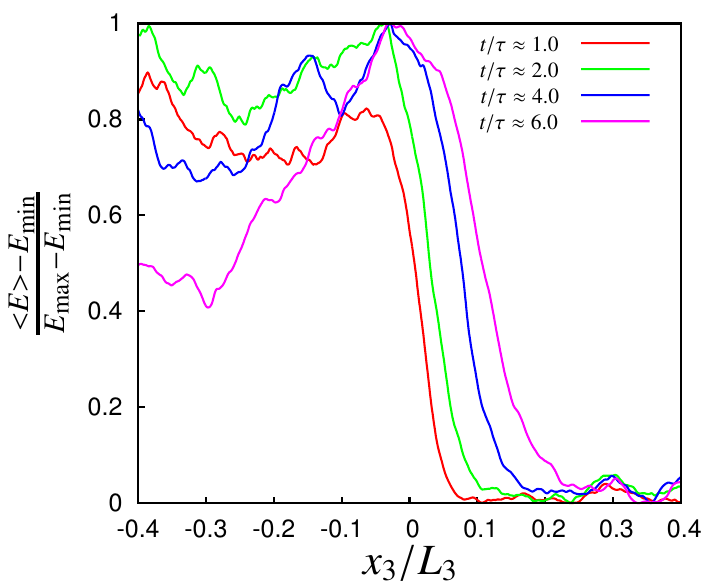}
		\caption{}
	\end{subfigure}
	
	\caption{{Mixing layer thicknesses. (a) temperature, (b) passive scalar vapor. Distrbution of the normalized kinetic energy at different time instants for  \fr=4.2 (c) and  \fr=-4.2 (e). Time evolution of the pit width with \fr. $E_{min}$ (d), $E_{max}$, minimum and maximum kinetic energy inside the mixing layer. The clear air top region in panel (c) (right part of the plot) initially shows a value of around 0. The pit onset starts at around $t/\tau=2$, and it is clearly visible beyond  $t/\tau=4$, when the layer portion with normalized energy close to 0 is located in the 0 - 1 range of  $(x_3-x_c)/\delta$. Panel (e), in this case, the clear air low energy region always shows a value of around 0. Instead, as a consequence, a temporal reduction in the high energy cloudy region highlights the formation of a peak which remains in the very center of the mixing.}}
	\label{fig:thicknesses}
\end{figure}

\begin{figure}[bht!]
	\centering
	\includegraphics[width=0.7\textwidth]{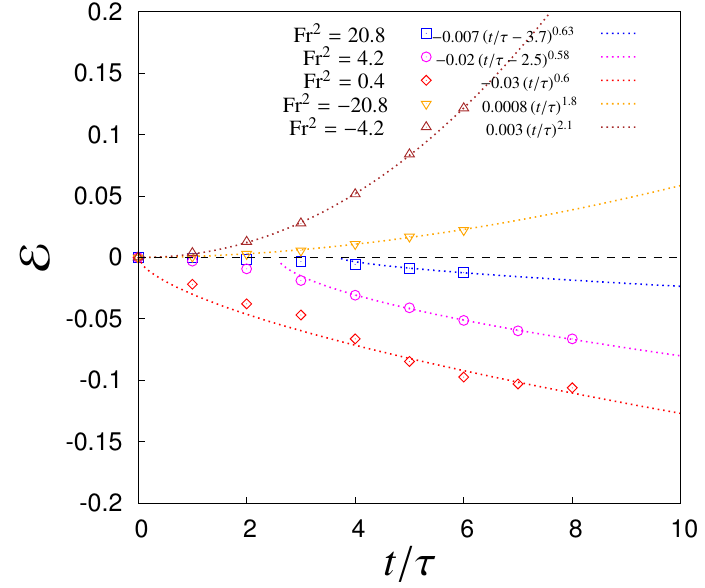}
	\caption{Time evolution of the relative turbulent energy variation $\mathcal{E}$, which is defined as the difference in the  kinetic energy inside the mixing layer from the neutral case \fr=69.2 (see equation \ref{eq:evar}).  $\mathcal{E}$ follows an algebraic trend. In  stable cases, after an initial transition that can last four eddy turnover times, a decay of the relative energy is observed inside the mixing,  with lower exponents than 1. In unstable cases, the exponents are greater than 1 and the initial transition is absent.}
	\label{fig:decgro} 
\end{figure}

In section \ref{sec:stat}, the onset of a sublayer can be observed beyond the time instant of the transient when buoyancy starts to be non-negligible in the center of the domain, where the initial temperature gradient is located. The formation and time evolution of such a sublayer are shown in Fig. \ref{fig:thicknesses}, where the time variation of the temperature and vapor interface thicknesses and the normalized kinetic energy profiles,   $E_\mathrm{norm}=\left(\langle E \rangle - E_\mathrm{min}\right)/\left(E_\mathrm{max}-E_\mathrm{min}\right)$, are shown. Here, $E_\mathrm{max}$  and $E_\mathrm{min}$ are  the maximum and the minimum mean kinetic energies, respectively. The normalized energy is almost equal to 0 in the low energy clear-air region, and nearly equal to 1 in the high energy cloud vapor region. In stable cases, the presence of the pit of energy changes the location of $E_\mathrm{min}$, which is now placed inside the pit, while $E_\mathrm{max}$  always remains inside the high energy region. As a consequence, after the onset of the pit, $E_\mathrm{norm}$ is approximately equal to 1 in the high-energy region, to 0 inside the pit and to $>0$ in the low energy region, as can be observed in panel (c) in Fig. \ref{fig:thicknesses} for the \fr=4.2 case. An opposite trend can be observed in unstable cases, after the formation of their peak sublayer: $E_\mathrm{norm}$ is 0 in the low energy region, 1 inside the peak sublayer, and $<1$ in the high energy region, see panel (e) in the same figure.

It is worth analyzing these "loss" or "gain" variations with reference to the neutral case of \fr=69.2. We can define the following relative kinetic energy variation: 
\begin{equation}
	\label{eq:evar}
	\mathcal{E}= \frac{E_{\mathrm{mix}}-E_{\mathrm{mix,Fr}^2=69.2}}{E_{\mathrm{mix}}+E_{\mathrm{mix,Fr}^2=69.2}},
\end{equation}
where $E_{\mathrm{mix}}$ and $E_{\mathrm{mix,Fr}^2=69.2}$ are  the kinetic energies within the mixing layer. This variation is obtained by integrating over thickness $\Delta_{\chi}$,  which is conveniently defined on the passive scalar distribution. In fact, the complex behavior of the kinetic energy profiles makes it difficult to provide an unambiguous definition of the layer thickness. The definition of $\Delta_{\chi}$ is given by
\begin{equation}
\label{eq:delta}
\delta_\chi(t) = x_\mathrm{top}(t) - x_\mathrm{bot}(t)
\end{equation}
\noindent where $x_\mathrm{top}$ and $x_\mathrm{bot}$ are the vertical locations in which the mean scalar concentrations are equal to 0.25 and 0.75, respectively:
%such that, considering $\Delta E(t)$ the mean difference of kinetic energy between the two external regions,
$$
{\langle \chi\rangle (x_\mathrm{top},t)} = 0.25
\qquad\qquad
{\langle \chi\rangle (x_\mathrm{bot},t)} = 0.75,
$$
see  Fig. \ref{fig:thicknesses}(b). However, it should be noted that, in the absence of any stratification, the thicknesses of the algebraic growth of both the passive scalar and the kinetic energy have a common exponent, see Figure 6 in \cite{jturb} and also \citep{vw89, vw90}.

The time evolution of $\mathcal{E}$ is shown in Fig. \ref{fig:decgro}. The relative energy variation in the presence of  unstable stratification increases in time with an algebraic trend; the exponents increase as the stratification increases -- 1.84 for \fr=-20.8, 2.14 for \fr=-4.2. The situation is more complex in stable cases. An initial transition phase can be observed, where $\mathcal{E}$ is almost constant.  There is then an algebraic  decay, with lower exponents than 1.  It should be noted that the initial transition is not present in the case of a very strong stable stratification (\fr =0.4).

We define the pit sublayer as the region where the kinetic energy (averaged in the $x_1-x_2$ planes) is lower than $80\%$ of the mean energy inside the low energy clear region. The intensity variation of the energy pit sublayer in time is represented in Figure \ref{fig:thicknesses}(d).  After the initial transition, the pit width grows almost linearly in time. This is in good agreement with the hull length growth found in a stratified Rayleigh-Taylor instability simulation by \cite{bifer2011} and, at least qualitatively, with the temporal evolution of the downdraught  penetration length in bouyancy reversal in cloud tops \cite{mel2009}.

It can be seen, from Fig. \ref{fig:thicknesses} (a,b), that the thickness of the mixing layer is still growing during pit formation. 
Only after a couple of time scales beyond the pit onset does the growth stop, and it is then followed by small oscillations around an asymptotic value. A different behavior is observed for unstable stratifications. In these cases, the generation of the energy peak enhances the mixing by providing a faster thickening of the layer, with  greater exponents, that is, -- 0.63 for \fr=-4.2, 0.54-0.56 for \fr=-20.8 -- than the neutral case for which the exponent is 0.42-0.49.
 
\subsubsection{Transport and Entrainment}

\begin{figure}[bht!]
	\begin{minipage}{.5\textwidth}
		\centering
		\includegraphics[width=\textwidth]{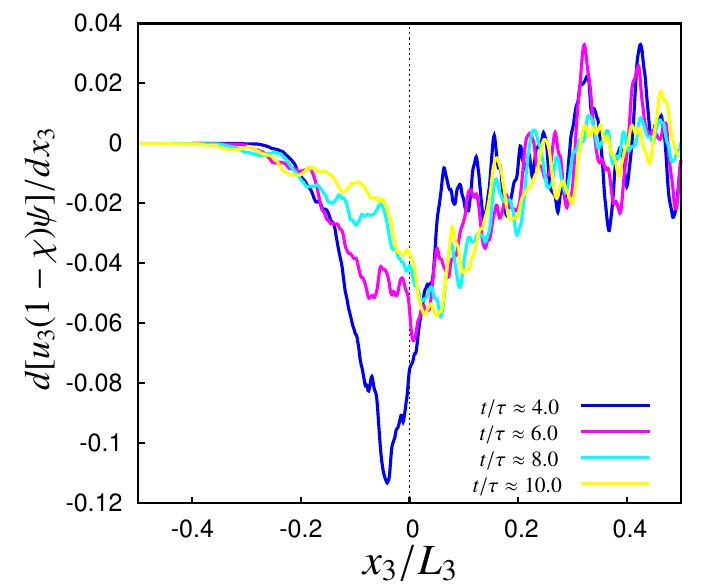}
		\hspace*{1cm}
		(\textit{a})\ $t/\tau = 4$, vapor flux at Fr $= 2.05$
	\end{minipage}\hfill
	\begin{minipage}{.49\textwidth}
		\centering
		\includegraphics[width=\textwidth]{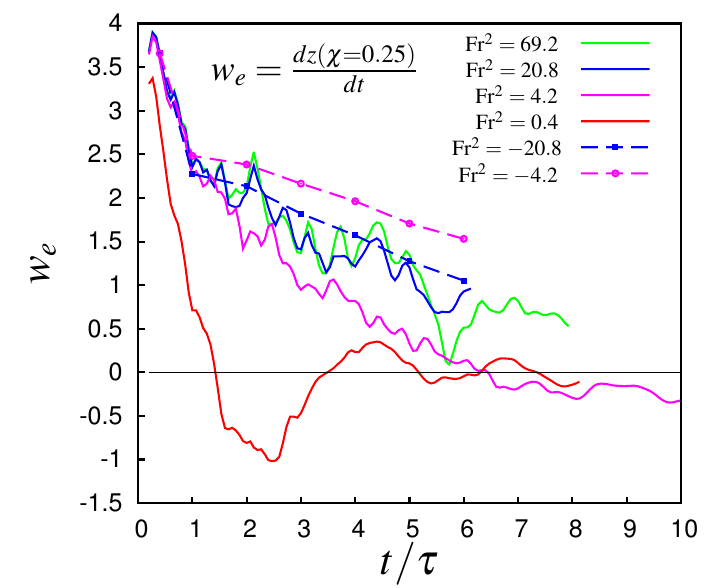}
		\hspace*{1cm}
		(\textit{b})\  Entrainment velocity
	\end{minipage}	
	\caption{Panel (\textit{a}): vertical variation of the mean flux of the vapor in the cloud; the marker function $\psi$ only takes into account the spatial points where the velocity is directed downward.
	Panel (\textit{b}): time evolution of the mean entrainment velocity fluctuation, $w_e$, which is normalized with the high kinetic energy $E_1$ root mean square. $w_e$ is calculated in the horizontal plane where $\chi=0.25$. Both stable interfaces (solid lines) and unstable interfaces (dotted lines) are represented here.}
	\label{fig:entrScal}
\end{figure}

{The entrainment of external fluid inside mixing layers is an important inertial aspect of interface dynamics, and, such an entrainment can range from those of the typical turbulent - non turbulent interfaces of boundary layers, jets, hyperbolic tangent shear layers, and wakes, to those of the shear-free interfaces observed in planet atmospheres and astrophysical clouds. }% \cite{wo12}.
Only downward velocity fluctuations can transport clear air into a vapor cloud in any plane parallel to the interface, in the absence of a mean velocity. Their presence can be highlighted  by a marker function, $\psi$, that is equal to 1 when $u_3$ is negative, and 0 otherwise. The entrainment mean value outside the mixing region is approximately constant and equal to $0.5\pm 0.01$, a value which would be observed for homogeneous and isotropic turbulence. Instead, the deviation inside the mixing layer is greater (up to $\sim\pm 0.05$), with a spatial distribution and a temporal evolution which somehow follow the ones observed in the third-order moment of the velocity, see Figure \ref{fig:skewVel}.
Figure \ref{fig:entrScal} (a) shows the vertical derivative of the downward vapor flux when Fr $= 2.05$.
The downward flux reduces as the flow evolves and its derivative, which represents the net variation of $1-\chi$ at a given instant, rapidly tends to zero inside the vapor cloud; this implies that the entrainment of clear air is confined to a thin interfacial layer.

{Since the entrainment of clear air is responsible for the growth of a cloud, it can be defined, and thus quantified, by considering the velocity with which the cloud expands. The velocity $w_e = dz/dt$, where $z = \langle x_{3,i}\rangle$ is the mean vertical position of the cloud top interface,  and  is here defined as the location where the mean vapor concentration $\chi$ is equal to 25\%. The time variation of $z$ has often been used as a parameter to measure the entrainment rate, see, for instance  \cite{mel2010,moe2000}.}

Figure \ref{fig:entrScal} (b) shows the time evolution of $w_e$ for different perturbation stratification levels. In the presence of a  quasi-neutral stratification, $w_e$ gradually decreases, with an algebraic trend, which is related to the natural decay of the turbulent kinetic energy. {\bf {However, when a stable, strong  stratification is present, the decay of $w_e$ is much faster and the entrainment vanishes after a few times scales. It should be noted that such an entrainment is related to the mixing thickness (see Fig. \ref{fig:thicknesses}), since the presence of the kinetic energy pit reduces the transport efficiency. On the other hand, in the case of unstable stratification, the presence of a kinetic energy peak enhances the mixing, and the entrainment speed therefore decays more slowly.}}

\begin{figure}[bht!]
	\begin{minipage}{.5\textwidth}
		\centering
		\includegraphics[width=\textwidth]{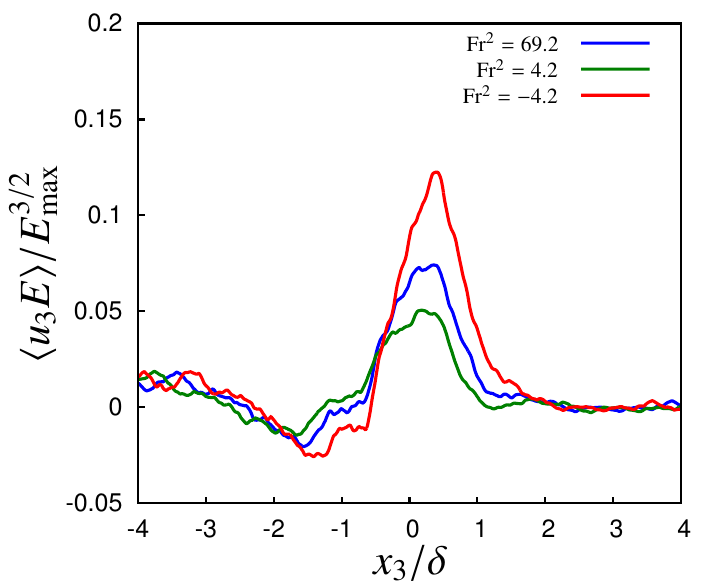}
		\hspace*{1cm}
		(\textit{a})\ $t/\tau = 3$, kinetic energy flux
		\includegraphics[width=\textwidth]{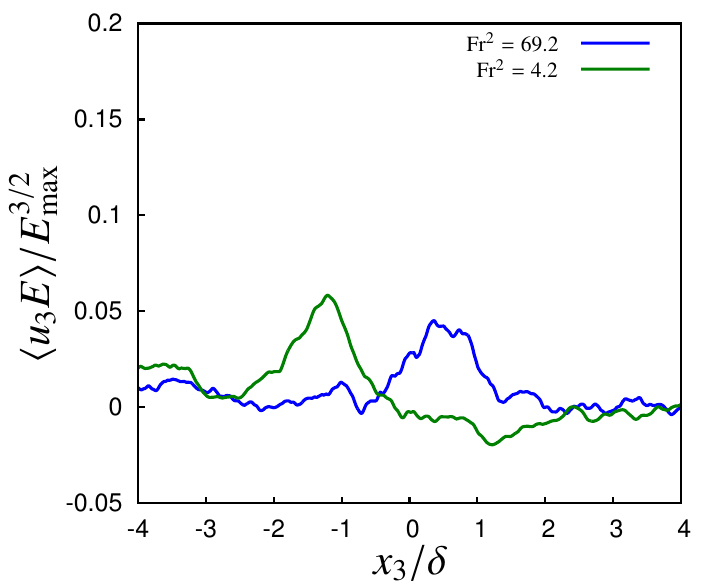}
		\hspace*{1cm}
		(\textit{c})\ $t/\tau = 8$, kinetic energy flux
	\end{minipage}\hfill
	\begin{minipage}{.5\textwidth}
		\centering		\includegraphics[width=\textwidth]{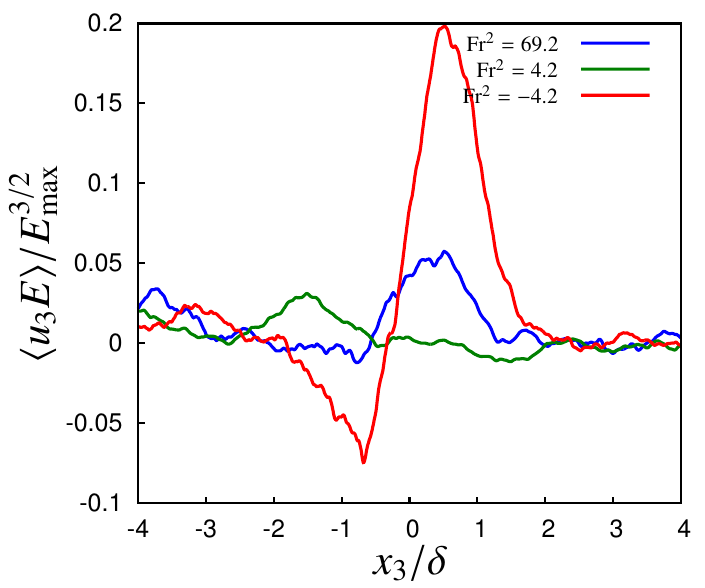}
		\hspace*{1cm}
		(\textit{b})\ $t/\tau = 6$, kinetic energy flux
		\includegraphics[width=\textwidth]{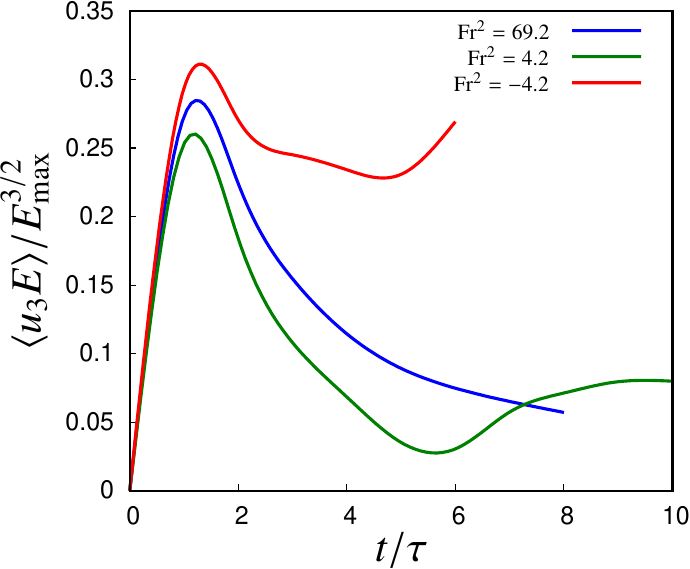}
		\hspace*{1cm}
		(\textit{d})\ Maximum of flux over time
	\end{minipage}	
	\caption{Panels (\textit{a--c}): kinetic energy fluxes along the vertical direction $x_3$, averages of the horizontal planes $x_1-x_2$, after 4, 6 and 8 time scales, respectively. It should be noted that, for the unstable simulations, it is not possible to reach 8 time scales for the computational stability problems self-generated by the physical condition of the flow.
	Data from simulations considering different initial squared Froude's numbers normalized on the mean kinetic energy of the high energy vapor cloudy region. Panel \textit{d} shows the temporal trend of the maximum normalized kinetic energy flux.}
	\label{fig:fluxVel} 
\end{figure}

\begin{figure}[bht!]
	\centering
	\begin{minipage}{.45\textwidth}
		\centering
		\includegraphics[width=\textwidth]{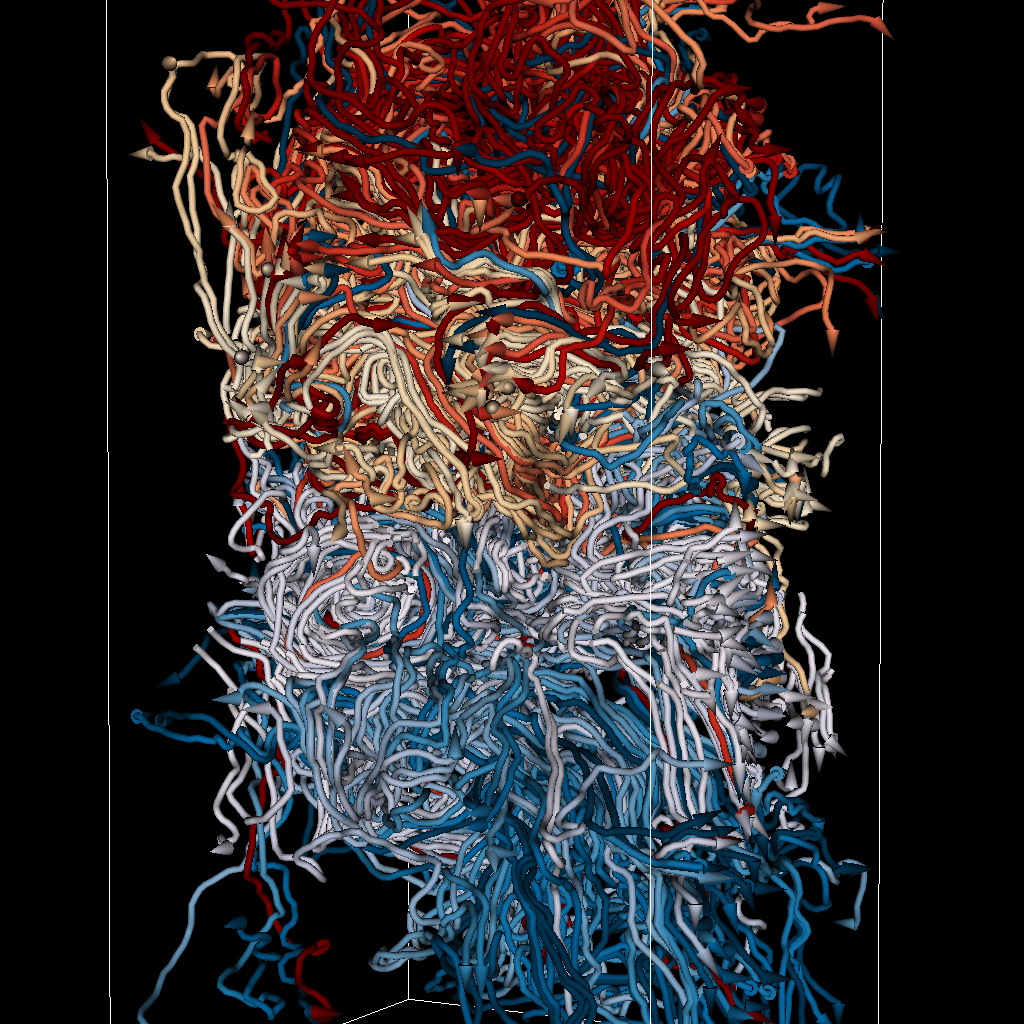}
		%		\hspace*{1cm}
		(\textit{a})\ \fr=4.2
	\end{minipage}\hfill
	\begin{minipage}{.45\textwidth}
		\centering
		\includegraphics[width=\textwidth]{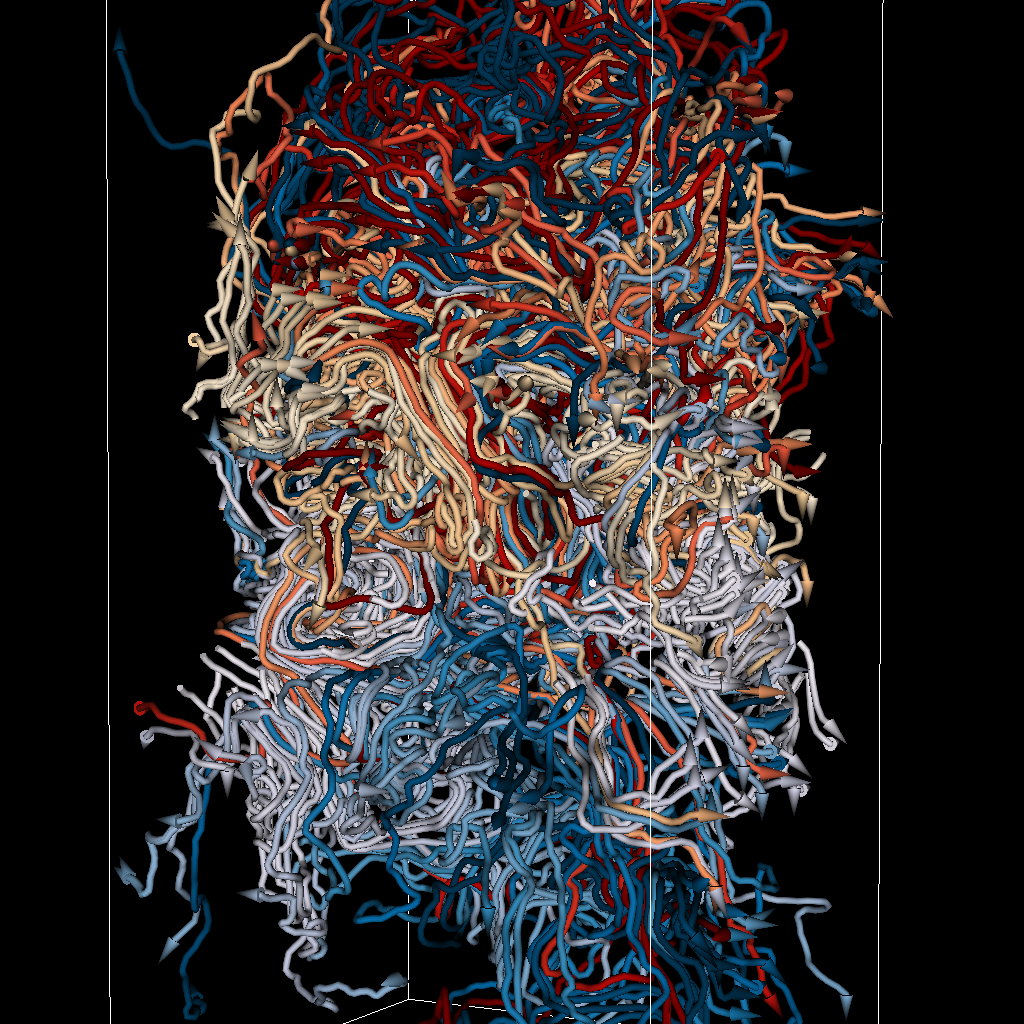}
		\hspace*{1cm}
		(\textit{b})\ \fr=-69.2
	\end{minipage}\\
	\vspace{.2cm}
	%	\begin{minipage}{.49\textwidth}
	\centering
	\includegraphics[width=.5\textwidth]{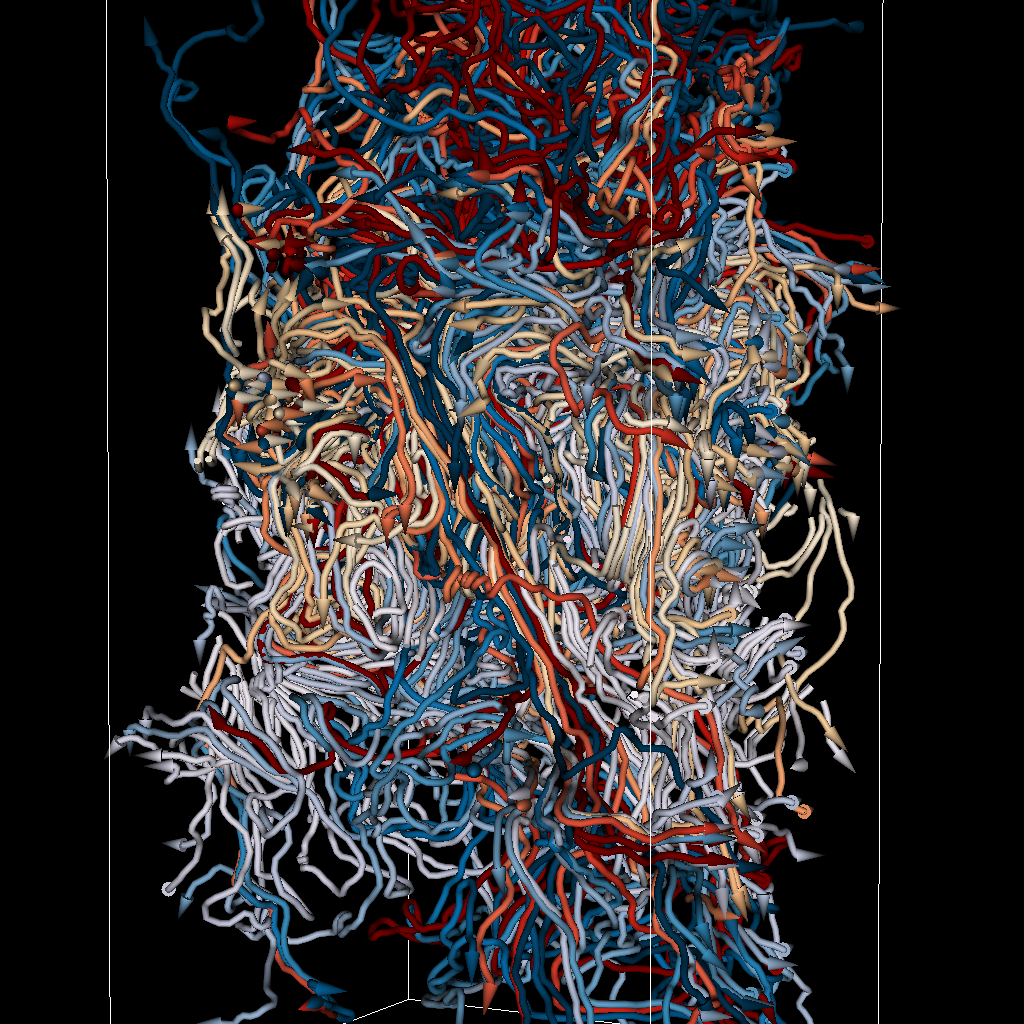}\\
	\hspace*{1cm}
	(\textit{c})\ \fr= - 4.2		
	%	\end{minipage}\hfill
	%	\begin{minipage}{.49\textwidth}
	\caption{Streamlines after 6 time scales for different stratification levels -- (\textit{a}) \fr=4.2 highly stable, (\textit{b}) \fr=69.2 negligible stratification, (\textit{c}) \fr=-4.2 highly unstable. The starting position of each streamline is placed at a fixed distance from above (yellow/red tubes) and below (cyan/blue tubes) the center of the interface. In panel (\textit{b}), where the buoyancy forces are negligible, streamlines from the upper side can cross the interface to reach the lower region, and viceversa. Instead, in panel (\textit{a}), where stable stratification effects are relevant, crossing of the interface becomes increasingly rare: what is located on one side of the interface tends to stay there, and the mixing process is damped. Finally, in the case of unstable stratification  shown in panel (\textit{c}), the mixture of red and blue lines is enhanced, which means that the streamlines cross the interface more frequently.}
	\label{fig:stream}
	%	\end{minipage}
\end{figure}

{The fact that a different level of entrainment is related to different efficiencies of the transport of any physical quantity can also be appreciated by observing the kinetic energy flux shown in Fig. \ref{fig:fluxVel}, and the passive scalar flux shown in Fig. \ref{fig:fluxes-active-passive-scalars}. Compared to a neutral case, the presence of stable or unstable stratification produces an initial reduction/increase in the energy flux, respectively, with a maximum flux always positioned around $x_3/\delta = 1$, see Fig. \ref{fig:fluxVel}. In the case of stable stratification, the flux decreases until it reaches a very low value. The formation of two fluxes can then be observed, according to the experimental results of \cite{jw94}. The first one, which is located below the pit at $x_3/\delta=-1$, is positive (upward flux), and the second one, which is located above the pit, in between  $x_3/\delta=1$ $3$, is negative (downward flow) -- see Figure \ref{fig:thicknesses}(c) and Figure \ref{fig:fluxVel}(c). A minimum value of 0.025 for the stable case with \fr$=4.2$ can be noted for the time evolution of the integral value of the flux in the layer, see, Figure \ref{fig:thicknesses}, panel (d). 
The ratio between the two fluxes is around 0.25 for \fr=4.2. No mean flux is present in between these two fluxes: this means that the energy tends to accumulate at the pit edge without being able to cross it, thus limiting the mixing thickness to a fixed width.
 In particular, if panel (a) in Figure 17 of JW is compared with the temporal sequence of panels a, b, c in our Fig. \ref{fig:fluxVel}, it can be seen that both show a reduction and inversion of the kinetic energy flow in the case of large stable stratification. The trend in panels (b, c) in Figure 17 in JW also shows a substantial agreement between the derivative of the energy flow along the vertical and the trend of our flow for \fr = 69.2 and 4.2 at the end of the transient. 
 In the case of unstable stratification, although the maximum flux located at $x_3/\delta=1$ keeps growing, the formation of a secondary negative flux, located near $x_3/\delta=-1$, can be observed. In this case, the energy is spread from the peak sublayer to the external homogeneous vapor cloudy region, thereby promoting mixing layer thickening.}

A similar behavior characterizes the passive scalar flux, which is shown in the top panel on the right in Figure \ref{fig:fluxes-active-passive-scalars}. The unstable stratification enhances the flux, which becomes increasingly important, in comparison with the scalar variance. However, no particular changes in the spatial trend can be seen: the flux is always directed toward the upper region. On the other hand, important differences can be seen for the case of stable stratification: after an initial damping, the flux becomes zero or even negative in the center of the mixing layer (see \fr=0.4 in the above cited panel), %figure \ref{fig:fluxScal}(\textit{a}, and see \fr=4.2 in figure \ref{fig:fluxScal}(\textit{c}), 
which agrees with the experimental results of \cite{jw94}. In particular, the flux derivative along the vertical direction can be noticed in the bottom panel on the right. %in Fig. \ref{fig:thicknesses}. %, shown in figure \ref{fig:fluxScal}(\textit{d}). 
A positive derivative can be interpreted as the entrainment of clear air (the passive scalar moves away), while a negative derivative implies a detrainment of clear air (the passive scalar moves into the layer) \cite{mw86}. In the case of neutral (and unstable) stratification, the mixing moves the scalar upward, where it is not initially present. We observe two sub-layers (dark yellow solid line, \fr$=0.4$) for a stable stratification with a positive derivative that surrounds one sub-layer with a negative derivative: the scalar is thus retained within the mixing layer.

A reduction in communication between the two regions external to the mixing layer can also be observed by looking at an instantaneous three-dimensional visualization of the flow streamlines, see Fig. \ref{fig:stream},  where  three stratification cases are shown: neutral \fr$=69.2$, stable \fr$=4.2$, and unstable \fr$=-4.2$. The streamlines are computed for fluid particles initially placed at a distance of $2\delta_0$ above (red) and below (blue) the center of the mixing layer, and are visualized at 6 initial eddy turnover times. It is possible to observe that, in the neutral case, panel (b), streamlines from the upper side can cross the interface to reach the bottom  region, and viceversa. This does not happen in the presence of stable stratification, panel (a); in this case, crossing of the interface becomes increasingly rare, and almost all the particles  located on one side of the interface remain there. On the contrary, in the presence of unstable stratification, panel (c),  mixing is enhanced and the streamlines cross the layer more frequently.

\subsubsection{Anisotropy and dissipation}\label{sec:anis}

\begin{figure}[bht!]
	\begin{minipage}{.5\textwidth}
		\centering
		\includegraphics[width=\textwidth]{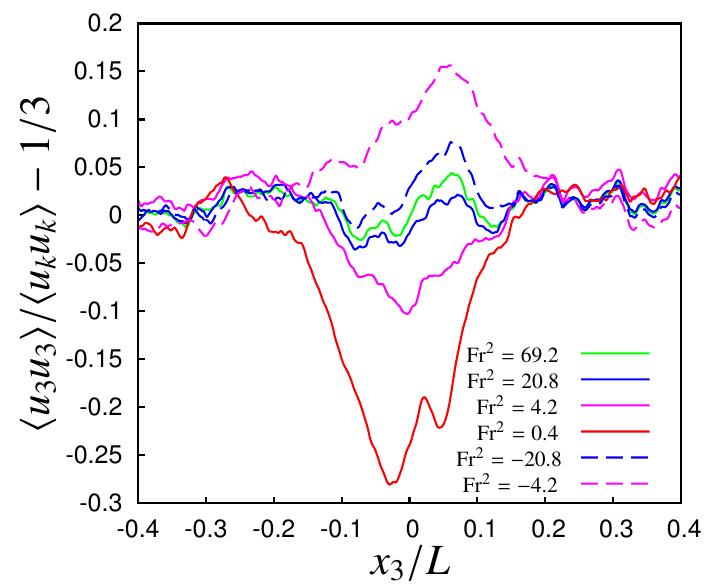}
		\hspace*{.5cm}
		(\textit{a})
	\end{minipage}\hfill
	\begin{minipage}{.5\textwidth}\centering
		\includegraphics[width=\textwidth]{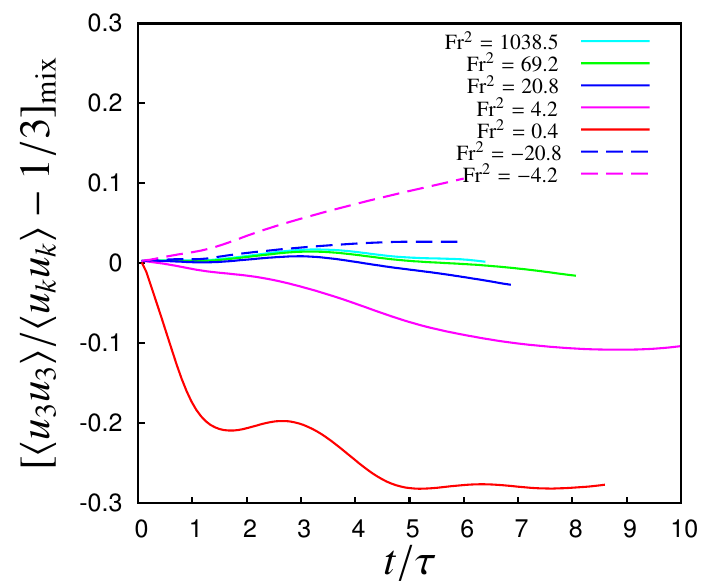}
		\hspace*{.5cm}
		(\textit{b})
	\end{minipage}
	\caption{Anisotropy of the turbulence large scales. (\textit{a}) $B_3$ ratio along the vertical direction obtained by varying the \fr number. (\textit{b}) Temporal evolution of  the $B_3$ peak value.}
	\label{fig:skeConf}
\end{figure}

Since buoyancy forces act mainly on a vertical component of the velocity field, an anisotropy enhancement, with respect to the unstratified situation, can be expected for both the large-scale turbulence and the small-scale  turbulence. We consider the relative weight of the energy associated with the vertical velocity, with respect to the other components \citep{pope}, to evaluate the large-scale anisotropy, using the ratio
$$
B_3=\frac{\langle u_3 u_3\rangle}{\langle u_k u_k\rangle}-\frac{1}{3}.
$$
Figure \ref{fig:skeConf} shows the behavior of the $B_3$ ratio along the vertical direction (panel \textit{a}), and the time evolution of its peak value in time. Anisotropy is present in a large-scale for a neutral stratification condition, but is limited, with a maximum deviation of 5\%. Anisotropy becomes very intense in the presence of a stratification. It is in particular possible to observe that the vertical fluctuation undergoes a large dumping under stable stratification conditions ($\langle u_3^2\rangle < \langle u_{1,2}^2\rangle$) and, viceversa,  an intense growth under unstable conditions ($\langle u_3^2\rangle > \langle u_{1,2}^2\rangle$). These differences are responsible for the different behavior of the transport and fluxes observed in the previous sections. 
It is also possible to observe, in Figure \ref{fig:skeConf}\textit{a}, that the variation concerns the global mixing layer (and not only the previously introduced pit/peak sub-layers of the kinetic energy). In fact, together with the formation of such sub-layers, a concomitant shift in the main energy gradient is also observed. This fact confirms the observation that the time evolution of a cloud during mixing is somewhat sensitive to large-scales \cite{Gotzfried2017} in concomitance with the important effects induced directly on the drop size distribution and supersaturation fluctuation by the small-scale \cite{golshan2021}.

As for the small-scale anisotropy of the flow, it should be mentioned that it is accurately represented by the higher moments of the first-order longitudinal derivative of the velocity components \citep{sa97}. It is well known that HIT departs from Gaussianity at small scales, and the longitudinal derivative skewness, $S(\partial u_i/\partial x_i) $, is almost equal to $ -0.5\pm0.1$, with a slight dependency on the Reynolds number \citep{sa97}. In previous works \cite{tib08,prl11}, it was found that, in the presence of a mixing layer due to a mean kinetic energy gradient, at Taylor Reynolds’ numbers of between 45 and 150, $S(\partial u_i/\partial x_i)$ not only shows that there is a significant departure of the longitudinal velocity derivative moments from the values found in homogeneous and isotropic turbulence, but also that the variation in skewness has the opposite sign for the components across the mixing layer and parallel to it.
The anisotropy induced by the presence of a kinetic energy gradient also has a very different pattern from the one generated by homogeneous shear. The transversal derivative moments in the mixing are in fact found to be very small, which highlights that the smallness of the transversal moments is not a sufficient condition for isotropy. In addition to the Reynolds number, the level of anisotropy depends on the energy gradient, see \cite{prl11}.

\begin{figure}[bht!]
	\begin{minipage}{.5\textwidth}
		\centering
		\includegraphics[width=\textwidth]{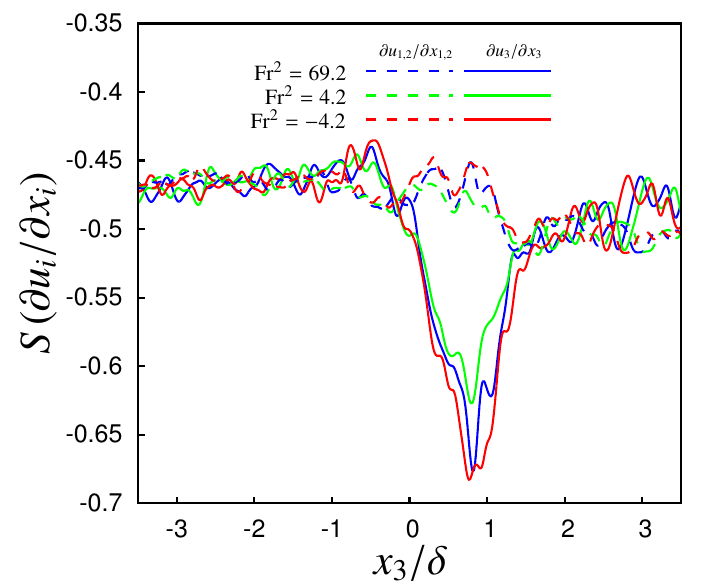}
		\hspace*{.5cm}
		(\textit{a})
	\end{minipage}\hfill
	\begin{minipage}{.5\textwidth}\centering
		\includegraphics[width=\textwidth]{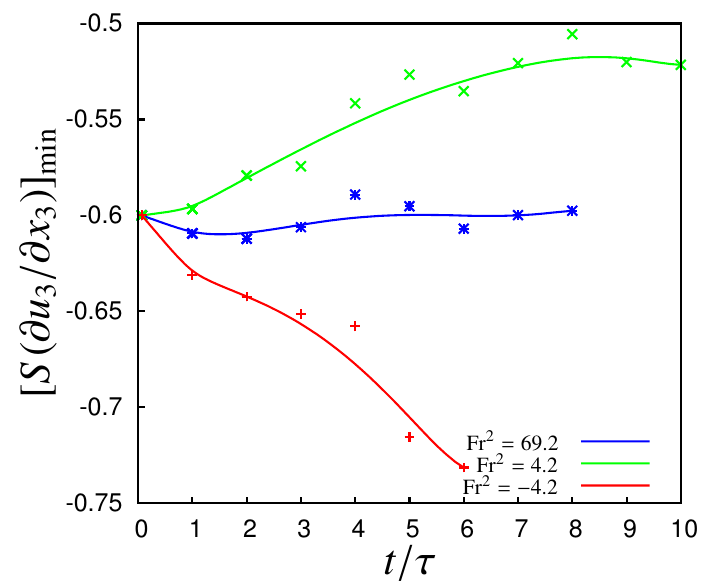}
		\hspace*{.5cm}
		(\textit{b})
	\end{minipage}
	\caption{Anisotropy of the Skewness of the longitudinal derivatives. (\textit{a}) Spatial distribution of the skewness of the longitudinal derivatives normal to the mixing surface (solid lines) and parallel to the mixing interface (dashed line). (\textit{b}) Evolution of the mean peak value of the longitudinal derivative crosswise direction of the mixing layer, the spatial location is  close to $x_3/\delta\approx1$. The symbols represent discrete computations, while the solid lines represent their spline interpolations.}
	\label{fig:skeDer}
\end{figure}

\begin{figure}[bht!]
	\begin{minipage}{.5\textwidth}
		\includegraphics[width=\textwidth]{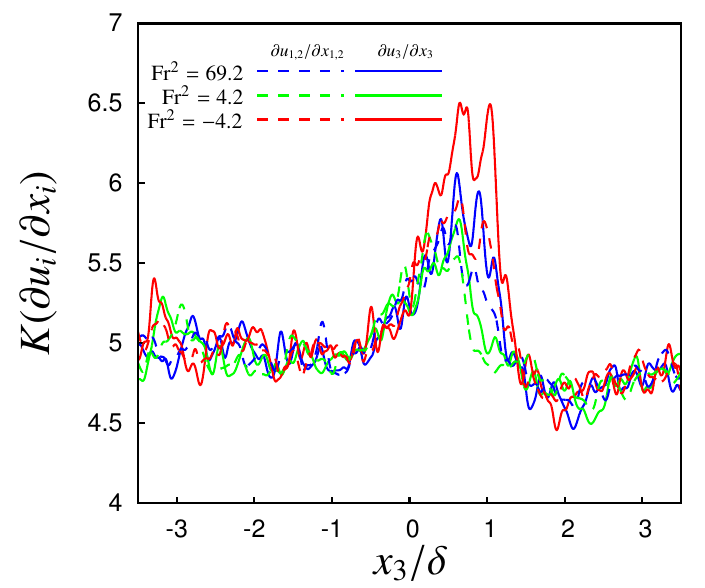}
	\end{minipage}\hfill
	\begin{minipage}{.5\textwidth}
		\includegraphics[width=\textwidth]{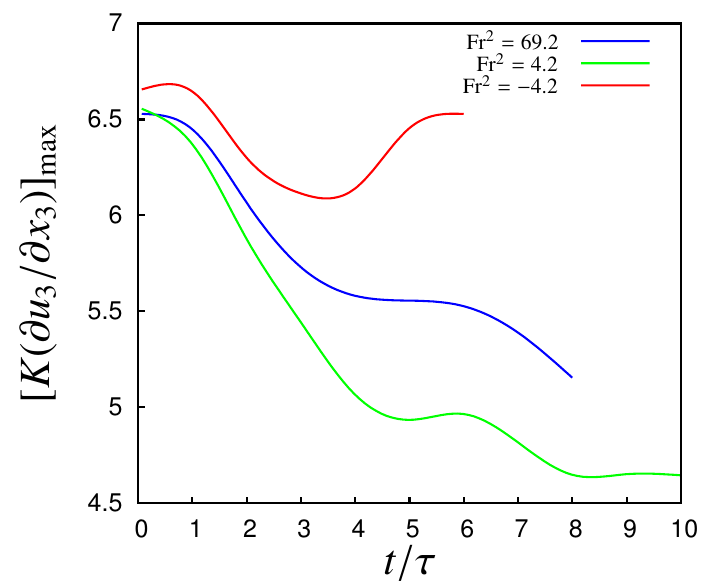}
	\end{minipage}
	\caption{Anisotropy of the Kurtosis of the longitudinal derivatives. (\textit{a}) Spatial distribution of the Kurtosis of the longitudinal derivatives normal to the mixing surface (solid lines) and parallel to the mixing interface (dashed line). (\textit{b}) Evolution of the mean peak value of the longitudinal derivative crosswise direction of the mixing layer, the spatial location is  close to $x_3/\delta\approx1$.}
	\label{fig:kurDer} 
\end{figure}

The presence of buoyancy forces does not directly influence the tilting/stretching of the vortex filament. Let us  consider the vorticity equation, obtained as the curl of equation \ref{eq:mom}:

\begin{equation}
	\frac{\partial\boldsymbol{\omega}}{\partial t} +\left(\boldsymbol{u}\bcdot\bnabla\right)\boldsymbol{\omega} = \left(\boldsymbol{\omega}\bcdot\bnabla\right)\boldsymbol{u} + \left(\boldsymbol{u}\bcdot\bnabla\right)+\nu\bnabla \boldsymbol{\times} \nabla^2 \boldsymbol{u} +  \alpha \bnabla \times (\boldsymbol{g}\theta),
	\label{eq:vort}
\end{equation}
where the compressibility stretching and the baroclinic terms have been neglected as a result of the incompressibility and Boussinesq's approximations. 
By considering that the buoyancy term is a vector that lies along the vertical direction, and its curl has horizontal  components which depend on the derivative along the direction parallel to the mixing:
\begin{equation}
	\alpha \bnabla \times (\boldsymbol{g}\theta) = \alpha \bnabla \times \left( \begin{array}{c}0\\0\\g\theta\end{array}\right) = \alpha g \left( \begin{array}{c}\partial\theta/\partial x_2\\\partial\theta/\partial x_1\\0\end{array}\right).
\end{equation}
it is possible to see that, if a mean variation of $\theta$ only occurs along the vertical, there will not be a mean contribution of buoyancy to the vorticity balance. Thus, in HIT, the presence of stratification does not influence small-scale anisotropy.  
However, this is not true inside a mixing layer. %: in particular, it has been found that a shear-less mixing present anisotropy at small-scales, with $S(\partial u_3/\partial x_3) < S(\partial u_{1,2}/\partial x_{1,2})$ inside the mixing region, with peaks toward the lower energy region (at $(x_3-x_c) / \delta\approx 1$).
 Figures \ref{fig:skeDer} and \ref{fig:kurDer} show that the presence of stratification modifies the behavior of the skewness and kurtosis of the longitudinal derivatives. Mild stratification does not affect small-scale anisotropy: the skewness of $\partial u_3/\partial x_3$ tends to an asymptotic value of $-0.63\pm0.02$, as expected, for Re$_\lambda\approx 200\div250$ \citep{prl11, shen2000}. In the case of stable stratification, the skewness of all the longitudinal derivatives tends to the isotropic value of $0.52$, while it tends to diverge in the case of unstable stratification $S(\partial u_3/\partial x_3)$, reaching values as low as $-0.75$ at \fr $= -4.2$, with an overgrowth of $30\%$.

%Those variations are related to the influence on mixing intensity due to the stratification. 
In the case of stable stratification, %the mixing can form an anisotropy sublayer, albeit slightly dumped with respect to the neutral case, in the very first time-scale. 
as soon as the energy pit appears, the mixing process  decreases and the small-scale anisotropy sublayer tends to disappear, as can be seen in Figure \ref{fig:skeDer}(\textit{b}), since the longitudinal derivative in the direction across the mixing is gradually reaching the typical value of homogeneous isotropic fields. Thus, the behavior of the system is similar to when the energy gradient is not present -- which would seem to indicate that the exchange of information between the two outer regions is blocked. 
On the contrary, mixing is enhanced in the case of unstable stratification, and the layer becomes even more anisotropic for small-scales, and acts as if the energy gradient is larger.
According to the results of \cite{prl11}, derivatives along a homogeneous direction do not show peaks in the center of the mixing layer, and it should be recalled that Re$_\lambda$ is 250 in the present cases.

%The statistical behavior of the derivative normal to the mixing layer, i.e. $\partial u_3/\partial x_3$, shows the presence of a peak inside the mixing layer, as shown in figure \ref{fig:kurDer}.

%\begin{figure}
%	\begin{minipage}{.5\textwidth}
%		\includegraphics[width=\textwidth]{Figure8a_spectraVelHigh}
%		\hspace*{1cm}
%		(\textit{a})\  High energy region ($x_3/L=-0.50$)
%	\end{minipage}\hfill
%	\begin{minipage}{.5\textwidth}
%		\centering
%		\includegraphics[width=\textwidth]{Figure8b_spectraVelMix}
%		\hspace*{1cm}
%		(\textit{b})\  Mixing region ($x_3/L=0.00$)
%	\end{minipage}	\centering{\includegraphics[width=.6\textwidth]{esponenteVelTimeAll}}
%	\caption{\label{fig:spectraVel} Panels (\textit{a--b}) Kinetic energy spectra at different instants for the case with \fr=4.2. Spectra are compensateded according to the Obukhov-Corrsin normalization, in which $E(\kappa)=0.4\epsilon^{-2/3}\kappa^{-5/3}$. Panel (\textit{a}) shows the spectra in the high energy region, while panel (\textit{b}) in the centre of the mixing layer.
%		Panel (\textit{c}) Evaluation of the spectra exponent in the inertial range at different position along the vertical direction $x_3$. Three different stratification are represented: unstratified case (\fr=69.2, black), stably stratified (\fr=4.2, green, for clarity shifted downwards by 0.3) and unstably stratified (\fr=-4.2, red, for clarity shifted upwards by 0.3). The dotted lines represents the standard -5/3 Kolmogorov slope. Error bars represent the uncertainity of the linear regression with which we estimated the exponent in a log-log plane.}
	%	\end{minipage}
%\end{figure}
 
%\subsubsection{Spectra}
\begin{figure}[bht!]
	\begin{minipage}{.5\textwidth}\centering
		\includegraphics[width=\textwidth]{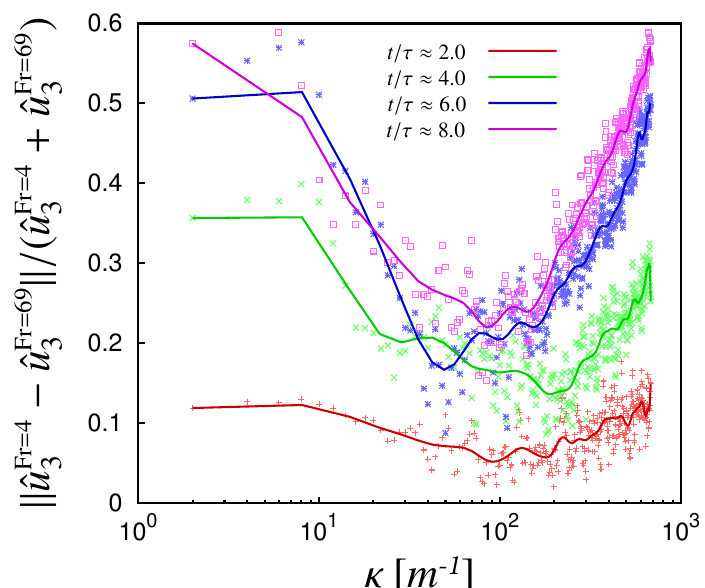}
		\hspace*{.5cm}
		(\textit{a})
		\includegraphics[width=\textwidth]{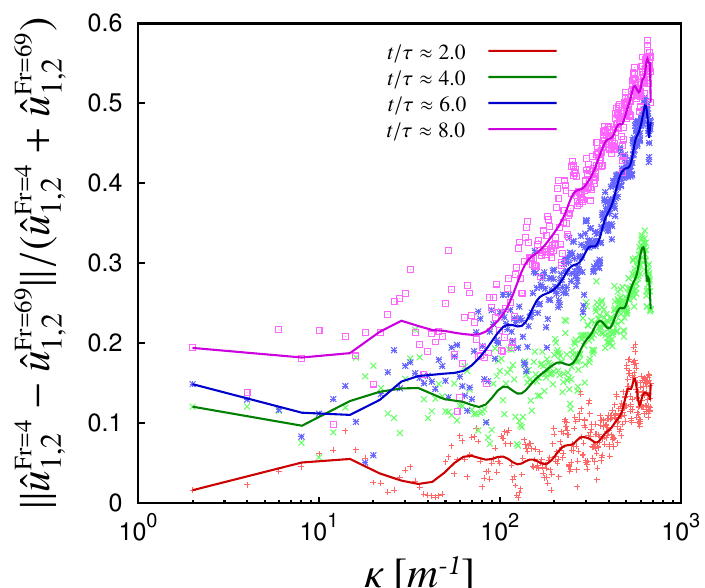}
		\hspace*{.5cm}
		(\textit{c})
	\end{minipage}\hfill
	\begin{minipage}{.5\textwidth}\centering
		\includegraphics[width=\textwidth]{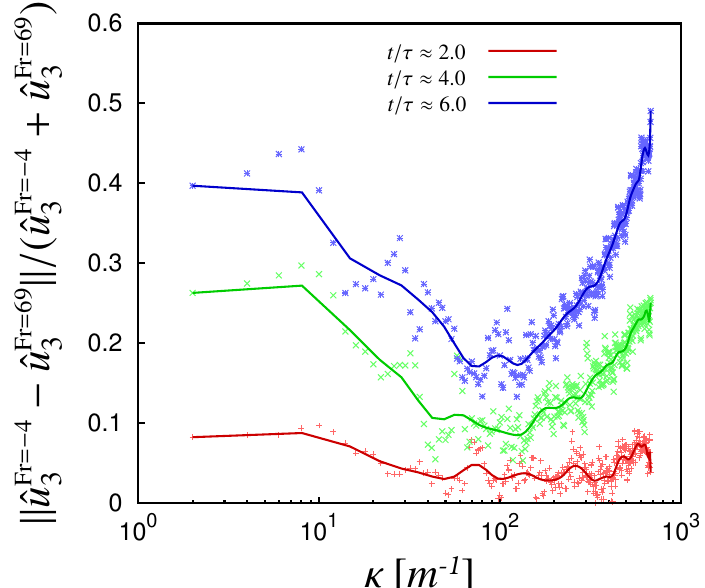}
		\hspace*{.5cm}
		(\textit{b})
		\includegraphics[width=\textwidth]{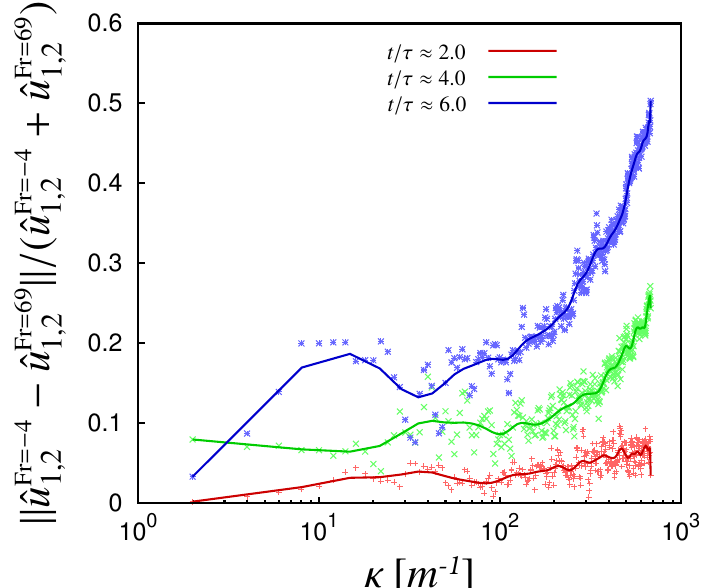}
		\hspace*{.5cm}
		(\textit{d})
	\end{minipage}
	\caption{ One dimensional velocity spectra along homogeneous directions.  Unlike the neutral case, \fr=69.2, panels (\textit{a--b}) show the effects on the vertical velocity fluctuations in the presence of stable (\textit{a}, \fr=4.2) and unstable (\textit{b}, \fr=-4.2) stratifications, respectively. Panels (\textit{c--d}) show the effects on the other two velocity components. The spectra were computed inside the mixing layer, at $x_3/\delta \approx 0.8$. Here, $\hat{u}_{1,2}$ is the arithmetic average of the one-dimensional spectra computed along the directions parallel to the mixing layer. The symbols represent the computation of discrete spectra, while the solid lines represent their B\'ezier interpolation.}
	\label{fig:confrSpec}
\end{figure}

%%%%%%%%%%%% Here I am!!!!!!

%We call $\tilde{E}(\kappa)$ the normalized Kolmogorov spectra of the  turbulent velocity:
%\begin{equation}
%	\widetilde{E}(\kappa) = 2.5 E(\kappa)\varepsilon^{-2/3}\kappa^{5/3}
%\end{equation}

%We synthesize the spectral exponent analysis in figure \ref{fig:spectraVel}\textit{c}, where the exponents of the inertial range are reported in function of time  for three different stratification level (stable, unstable and neutral); the values in case of stable and unstable stratification are displaced by $\pm0.3$ for clarity reasons. Exponents are estimated taking advantage of the linear regression method, applied to the spectra in a log-log space. The inertial ranges are chosen minimizing the regression error, under the condition to have at least a half decade in each inertial range. The error-bars in figure \ref{fig:spectraVel}\textit{c} represent the residual.

%Independently from the location and from the stratification effects, velocity spectra get in time close  to the Kolmogorov $-5/3$ slope.  The decrease of $\alpha$ exponent for the stably stratified case at $t/\tau>8$ is given by the absence of a neat logarithmic scaling, since the inertial range progressively shrinks in time because of the widening of the dissipative range, as previously shown in Fig. \ref{fig:spectraVel} (\textit{a-b}). 

Another interesting feature concerning anisotropy can be noted by observing the spectra at the edges of the inertial range in the presence of  stratification. This feature can be evaluated by comparing the one-dimensional spectra of each velocity component inside the mixing region with the neutral case. To achieve this, we computed spectra $\hat{u}_i(k,x_3)$ as the average of the transforms along each of the two homogeneous directions, that is 
\begin{equation}
	\hat{u}_i (k,x_3) = 0.5\langle\hat{u}_i(k_1,x_2,x_3)\rangle + 0.5\langle\hat{u}_i(x_1,k_2,x_3)\rangle\qquad k=k_1=k_2
	\label{spectra}
\end{equation}
where the average operator $\langle\cdot\rangle$ acts along the homogneous direction on which the transform is not carried out. The obtained spectra are then compared with the neutral case, \fr=69.2, by considering the relative variation 
$$
\frac{||\hat{u}_i^{\mathrm{Fr}^2=\dots}-\hat{u}_i^{\mathrm{Fr}^2=69.2}||}{||\hat{u}_i^{\mathrm{Fr}^2=\dots}+\hat{u}_i^{\mathrm{Fr}^2=69.2}||}.
$$
The results of such a comparison are shown in Figure \ref{fig:confrSpec} for the stable case, \fr=4.2 (panels \textit{a,c}), and the unstable case, \fr=-4.2 (panels \textit{b,d}). The first observation that can be made concerns the different behavior of the vertical velocity fluctuation from the other two components. As can be seen, stratification acts directly on the larger scale of the vertical motion, generating a relative deviation from the neutral case. Such a variation is negative (less energy in vertical motion) in the presence of a stable stratification, and positive (more energy) in unstable situations, in agreement with what has been observed for large-scale anisotropy. As the mixing evolves, these effects are transmitted to smaller scales through the inertial cascade, until the dissipative range is reached, with the consequent effect of enhancing/dampening of the dissipation rate for stable/unstable stratification, respectively. The stratification effects in this scale range are widespread in all the velocity components: as a consequence, absolute small-scale differences (and therefore small-scale anisotropy) are dumped in the presence of stable stratification, and enhanced in unstable cases.% as pointed out in section \ref{sec:anis}, see also Fig.s \ref{fig:skeDer} and \ref{fig:kurDer}.

We observed that Kolmogorov -5/3 scaling is present over the whole domain. The inertial range is rather narrow, as it extends for about one decade. % at $t/\tau>6$, see Fig. \ref{fig:spectraVel}\textit{a, b}. 
The normalized kinetic energy spectra are somewhat similar along the vertical direction, with small deviations, due to the different local Reynolds numbers. These spectra are also quasi-self similar in time, and the main difference is represented by a reduction in the extension of the inertial range due to the temporal growth of the Kolmogorov scale and of the dissipative range. A symmetrical variation has been observed in the inertial range, with respect to the non-stratified condition, on the spectral indices of the velocity spectra: -1.99, when \fr$= 4.2$, and -1.35, when \fr$= -4.2$.  The inertial range of the passive scalar power spectra shows an index of about -1.45 inside the cloud portion and of about -1.56 inside the mixing layer. These values slowly decrease over time.

\begin{figure}[bht!]
	\centering
	\begin{minipage}{.49\textwidth}
		\centering
		\includegraphics[width=\textwidth]{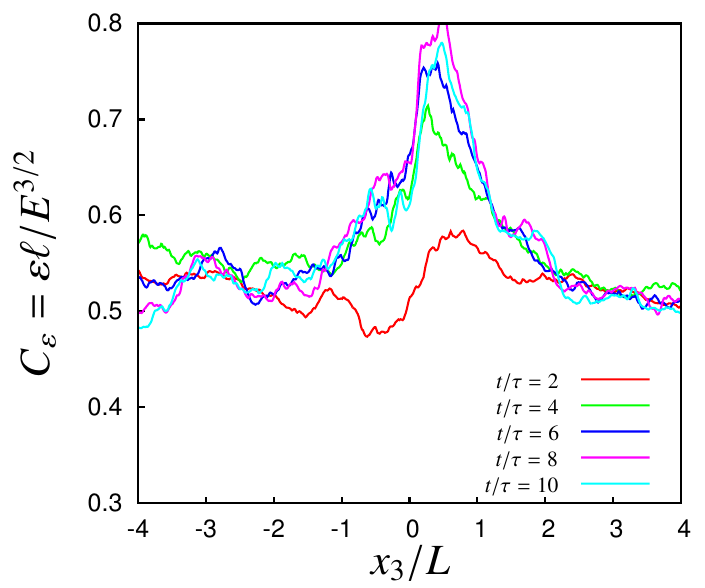}
		\hspace*{.5cm}
		(\textit{a})\ Normalized turbulent dissipation rate in a stable condition
		\end{minipage}\hfill
	\begin{minipage}{.49\textwidth}
		\centering
			\includegraphics[width=\textwidth]{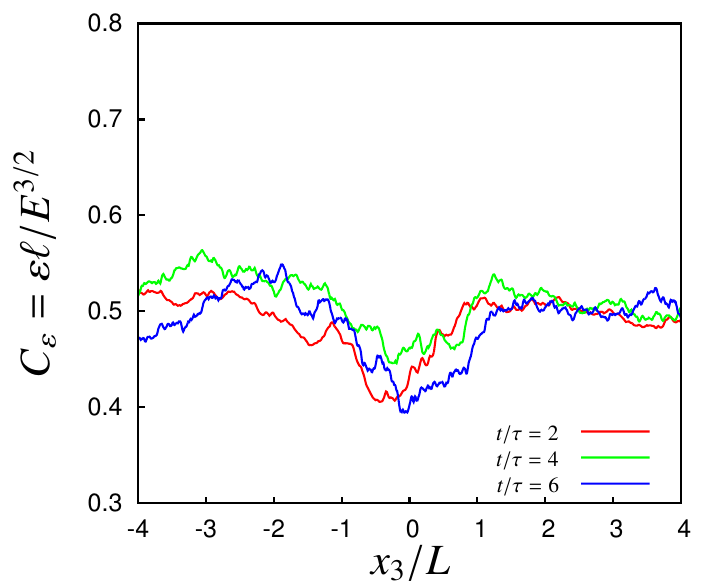}
		\hspace*{.5cm}
		(\textit{a})\ Normalized turbulent dissipation rate in an unstable condition.
	%	\includegraphics[width=\textwidth]{Figure7b_normDissPdf}		
	%	\hspace*{1cm}
	%	(\textit{b})\ Dissipation PDF's
	%	\includegraphics[width=\textwidth]{Figure7b_normDissPdfu}		
	%	\hspace*{1cm}
	%	(\textit{b})\ Dissipation PDF's
	\end{minipage}
	\caption{(\textit{a,b}) Normalized plane-averaged  dissipation $C_\varepsilon=\lbrace\epsilon\rbrace\ell/\lbrace E \rbrace^{3/2}$  for \fr=4.2 and \fr=-4.2, respectively. Values outside the mixing layer are close to 0.5, as in the case of isotropic homogeneous turbulence \cite{burat2005}. In stable cases, the normalized dissipation shows a maximum at the kinetic energy pit; in an unstable case, the dissipation shows a minimum at the energy peak.}
	\label{fig:normDiss}
	\end{figure}

The dissipation rate is computed over the whole domain using the general definition \citep[see ][p.64]{tenn}
\begin{equation}
	\varepsilon =   \frac{1}{2}\nu\left( \frac{\partial u_i}{\partial x_j}+\frac{\partial u_j}{\partial x_i}\right)^2.
\end{equation}
Figure \ref{fig:normDiss} shows the plot of the normalized turbulent dissipation rate $C_\varepsilon$, which is defined as
$$
C_\varepsilon (x_3)=\frac{\langle\varepsilon\rangle\langle\ell\rangle}{\langle E \rangle^{3/2}}
$$
where the averages in the horizontal planes have been implemented.
It can be observed that the normalized dissipation rate is initially almost constant in the transient and equal to $0.55\pm0.05$, that is, the same value as the unstratified case. %Actually there are not main differences also in comparison with an homogeneous isotropic turbulence, as in that case $C_\varepsilon$ tents to 0.5 for sufficiently high values of $Re_\lambda$ \cite{burat2005}.
However, as the buoyancy becomes relevant, in the case of stable stratification, \fr$=4.2$, the formation of a dissipation "peak" can be observed inside the pit of kinetic energy, where $C_\varepsilon$ reaches values as high as $0.9$, that is, an increse of nearly 70\%. On the other hand, in the case of unstable stratification, \fr$=-4.2$, the dissipation decreases inside the sublayer by nearly 20\%. Thus, dissipation is affected to a great extent by buoyancy. %, but to a lesser extent. As a result, the normalized dissipation becomes relatively larger in such sub-layers. This can be explained by considering that the stratification affects mainly the energy associated to vortices and less their geometry.  This assumption is confirmed with the probability density functions reported in figure \ref{fig:normDiss}(b) which show how in different parts of the system the statistical behavior of the dissipative rate remains similar inside and outside the mixing sub-layers.

\section{Concluding remarks}
%\label{conStrat}

The evolution of a freely decaying, shearless, turbulent mixing layer hosting both air and a vapor phase, considered as a passive scalar phase, is obtained by coupling two homogeneous isotropic turbulent fields with different kinetic energies.  A large range of Froude numbers (\fr $\in [-20.8, 1038.5]$) has been studied to evaluate the changes that take place in the mixing dynamics, due to both stable and unstable temperature conditions.

Our numerical simulations have shown that both stable and unstable stratifications modify the dynamics and transport characteristics of a shearfree turbulent layer. First, the formation of a sub-layer inside the mixing region is observed: i) a pit of kinetic energy under a stable condition, a sort of intense decay overshoot that is characterized by a lower level of energy than the external regions; ii) the formation of a peak of kinetic energy under unstable stratification conditions, where the turbulent energy becomes higher than in the external regions (15$\%$ larger at \fr$ = - 4.2$). The temporal scaling law of the energy variation inside the mixing region has been quantified. The exponent depends on the stratification intensity. It reaches a value of 2.1 at \fr$= -4.2$, which is about four times larger than the exponent determined at \fr $= 4.2$. Stable stratification almost suppresses vertical motion, since any fluctuations within it are inhibited by buoyancy forces. In such a condition, an increased anisotropy is observed for the large-scale structures, compared to the neutral case. In fact, the energy associated with vertical fluctuations gradually becomes smaller than the other components. On the other hand, vertical fluctuations, under unstable conditions, amplify with respect to the horizontal components.

Turbulence diffusion becomes damped in the presence of a stable stratification, as do intermittency, kinetic energy, passive scalar transport and clear air entrainment. Entrainment almost vanishes when the Froude square number becomes lower than 1. A detrainment phase, lasting from 1.4 to 3.5 eddy turns over time, is observed at \fr$= 0.4$. On the other hand, unstable stratification enhances the mixing process.

The dissipation function increases to a great extent for stable perturbation conditions. An increase of 70$\%$ at \fr $= 4.2$ has been observed here. Conversely, at \fr  $ = - 4.2$, a decrease of 2$\%$ has been observed. Log-normal probability density functions of the dissipation rate have resulted to be self-similar inside different layers across the mixing. This is a result that can be explained by considering that stratification has more effect on the energy associated with the vortical structures than on their morphology.

As far as small-scale anisotropy is concerned, it has been found that the presence of unstable stratification increases the differences in the statistical behavior between the longitudinal velocity derivatives. As a consequence, the compression of the fluid filaments normal to the interface is greater, due to the increased mixing intensity. Since the mixing process tends to vanish in stable cases, small-scale anisotropy also vanishes.

We have collected spectral information. The main observation concerns the velocity fields. By comparing the stratified spectral behavior with the unstratified behavior of the velocity fields, we have noted a substantial diversification in time for both low and high wave numbers for the vertical velocity fluctuations. Instead, for the horizontal components of the velocity fluctuation, differentiation is only clearly visible at the smallest scales, that is, for the highest wave numbers. 

{Looking ahead, we would like to conduct a simulation campaign on domains of a similar size to the size considered in this work, but including the aqueous liquid phase and the related collision and coalescence phenomena of water droplets, as has recently been done, albeit at a much smaller domain scale than the one considered here (Golshan et al. 2021 \cite{golshan2021}, Fossa' et al. 2022 \citep{fossa2022}). In particular, we would like to observe a longer time window, that is, a time corresponding to almost one minute of a three-phase (gas, vapor, liquid) warm cloud instead of the few seconds of the present simulation.
\\
However, it should be considered that droplet clustering  introduces a further complexity to the structure of the clear air-cloud interface. In particular, the discontinuous distribution  of droplets and droplet clusters in space means that different cores will require a very uneven computational effort at each time step, and this cannot be a priori predicted. In such a situation, where a physical modeling is still under  evolution, it would be very difficult to force the code to a massive high level of parallelization. In fact, the shift from slab to pencil parallelization (which has already been achieved for the version of the code used in this work, where water droplets are not simulated) increases the time needed to exchange information between the cores by about 8 times.  This occurs  because the amount of information exchanged by two adjacent cores is not homogeneous inside the computational domain and furthermore, it is likely that non-adjacent cores would also need  to exchange information. Such a situation has a high probability of occurring over  short time intervals, such as those that are comparable with a single computational time step, because turbulence hosts long-term phenomena which can induce large droplet displacements, that is, droplet displacements to a domain portion in a core not adjacent to the core where the droplet departed from.}

\section{Acknowledgments}
We gratefully acknowledge PRACE (project n$^\circ$ RA07732011) for having granted us access to the computational resources of Curie, France at TGCC, and SCAI (PRACE Type C project, GA 730913) Fermi, Italy at CINECA.

We acknowledge HPC@POLITO, which is  an Academic Computing project within the Department of Control and Computer Engineering at the Politecnico di Torino (\href{http://www.hpc.polito.it}{HPC@POLITO}).

We also acknowledge funding from the Marie-Sklodowska Curie Actions (MSCA ITN ETN COMPLETE) under the European Union’s Horizon 2020 research and innovation program. Grant agreement no. 675675, http://www.complete-h2020network.eu.

\section*{Data availability}

The data that were used to establish the findings of this study are available from the corresponding author upon reasonable request.

%We acknowledge the CINECA award HP10CA7H4X, under the ISCRA initiative, for the availability of high performance computing resources and support. Computational resources were also provided by HPC@POLITO, a project of Academic Computing within the Department of Control and Computer Engineering at the Politecnico di Torino (https://hpc.polito.it).

\section*{References}
\bibliography{GallanaEtAl_POF22-AR-01008}

\end{document}